\documentclass[12pt]{iopart}

\usepackage{epsfig} 
\usepackage{color}

\newcommand{\beq}{\begin{equation}}
\newcommand{\eeq}{\end{equation}}
\newcommand{\beqa}{\begin{eqnarray}}
\newcommand{\eeqa}{\end{eqnarray}}
\newcommand{\om}{\Omega_m}

\newcommand{\dls}{d_{\rm lss}} 
\newcommand{\fon}{\textcolor{white}} 
\newcommand{\dddot}{\tdot} 

\newcommand{\citep}{\cite}
\newcommand{\citet}{\cite}
\def\la{\mathrel{\mathpalette\fun <}} 
\def\gs{\mathrel{\mathpalette\fun >}}
\def\fun#1#2{\lower3.6pt\vbox{\baselineskip0pt\lineskip.9pt
  \ialign{$\mathsurround=0pt#1\hfil##\hfil$\crcr#2\crcr\sim\crcr}}} 
\newcommand{\ls}{\mathrel{\raise0.27ex\hbox{$<$}\kern-0.70em \lower0.71ex\hbox{{
$\scriptstyle \sim$}}}} 

\begin{document} 

%\title{Mapping the cosmological expansion} 
\review[Mapping the cosmological expansion]{Mapping the 
cosmological expansion} 
\author{Eric V Linder}
%\affiliation{} 
\address{Berkeley Lab \& University of California, Berkeley, CA 94720, USA} 
\ead{evlinder@lbl.gov}

\begin{abstract} 
The ability to map the cosmological expansion has developed enormously, 
spurred by the turning point one decade ago of the discovery of cosmic 
acceleration.  The standard model of cosmology has shifted from a matter 
dominated, standard gravity, decelerating expansion to the present search 
for the origin of acceleration in the cosmic expansion.  We present a wide 
ranging review of the tools, challenges, and physical interpretations.  
The tools include direct 
measures of cosmic scales through Type Ia supernova luminosity distances, 
and angular distance scales of baryon acoustic oscillation and cosmic 
microwave background density perturbations, as well as indirect probes 
such as the effect of cosmic expansion on the growth of matter density 
fluctuations.  Accurate mapping of the expansion requires understanding 
of systematic uncertainties in both the measurements and the theoretical 
framework, but the result will give important clues to the nature of 
the physics behind accelerating expansion and to the fate of the 
universe. 
\end{abstract} 

%\pacs{95.36.+x, 98.80.-k} 

%\keywords{cosmology: observations --- cosmology: theory --- supernovae}
%\date{\today}

%\maketitle 

\tableofcontents

\section{Introduction} \label{sec:intro}

A century ago our picture of the cosmos was of a small, young, and 
static universe.  Today we have a far grander and richer universe to 
inhabit, one that carries information on the strongest and weakest 
forces in nature, whose history runs from singularities and densities 
and temperatures far beyond our terrestrial and laboratory access to 
the vacuum and temperatures near absolute zero.  Understanding our universe 
relies on a wide range of physics fields including thermodynamics, classical 
and quantum field theory, particle physics, and gravitation. 

Perhaps most amazing is that while our universe is huge, it is finite 
in well defined ways, and can be encompassed and comprehended.  The 
visible universe extends for nearly an equal number of orders of magnitude 
above the size of the Earth as the proton lies below.  The number 
of particles is of order $10^{80}$, large but not unbounded, and the 
age of the universe is some 14 billion years, only three times the 
age of the Earth.  And the universe is simple in ways we have no right 
to expect: it is essentially electrically neutral and its large scale 
dynamics is governed by gravity and no other forces of nature, it is 
nearly in thermal equilibrium for most of its history, and the geometry 
of space is maximally symmetric. 

A few characteristics hold where we might be tempted to say, as 
Alfonso X ``The Wise'' did in the 13th century: ``Had I been present 
at the creation of 
the world, I should have recommended something simpler''.   The universe 
has about a billion times more entropy per baryon than expected, and 
in a related sense has a greatly unequal ratio of matter to antimatter. 
The dynamics is neither kinetic energy dominated nor potential energy 
dominated but apparently perfectly balanced, giving the critical energy 
density and a flat spatial geometry.  But the most important cosmological 
discovery of the 20th century was that the maximal symmetry does not 
extend to spacetime; that is, we do not live in a steady state universe 
unchanging in time. 

Discovery of the cosmic expansion of space in the 1910s and 1920s and 
that the evolution arose from a hot, dense, early state called the Big 
Bang in the 1960s gave 
rise to modern cosmology as a exemplar of, window on, and laboratory for 
physics.  Ten years ago the discovery of the {\it acceleration\/} of that 
expansion revolutionized cosmology and a great array of overlapping 
fields of physics.  This article addresses our current knowledge of 
the cosmic expansion and our prospects for exploring it in detail.  
In particular, we focus on the recent epoch of acceleration 
and the astrophysical tools for mapping the cosmic expansion history 
to reveal the nature of new physics beyond our present standard model. 

In the remainder of this section we discuss how the expansion of our 
universe impacts fundamental questions of the origin and fate of the 
cosmos and everything in it, and its intimate relation with the nature 
of gravity, as a force itself and reflecting on unification with 
quantum theory.  We present the effects of acceleration in \S\ref{sec:eff}, 
but do not go into the root causes of it, which are highly speculative 
at this time; see, e.g., the focus issue on dark energy in \citet{grgissue}. 
Techniques for directly mapping the expansion appear in \S\ref{sec:dist}, 
while \S\ref{sec:gro} briefly mentions indirect effects of acceleration 
through the growth of structure in the universe.  Obtaining an accurate 
map is crucial to our understanding, and \S\ref{sec:sys} focuses on 
challenges imposed by systematic uncertainties within the theoretical 
interpretation and data analysis.  For future measurements these may be 
the key issues in advancing our knowledge.  Future prospects for mapping 
the cosmic expansion from the Big Bang to the final fate are outlined in 
\S\ref{sec:fut}, and \S\ref{sec:concl} summarizes and concludes.

\subsection{The dynamic universe} 

In the 1910s the frequency shift of spectral lines from astronomical 
sources with respect to laboratory measurements were observed by 
Slipher and the dynamics of space was found by Einstein within his 
theory of general relativity.  While Einstein, together with de Sitter, 
acted to counteract the expansion of space and retain a static state 
through introduction of 
the cosmological constant, other researchers in the 1920s such as Friedmann 
and Lema\^{\i}tre calculated the expansion history of a universe 
containing matter and components with pressure, and Weyl recognized that 
a physical expansion scaling linearly with distance occurred naturally.  
The observations of Hubble in 1929 established the expanding universe as 
the basis of cosmology. 

\begin{figure}[!htb]
\begin{center} 
\psfig{file=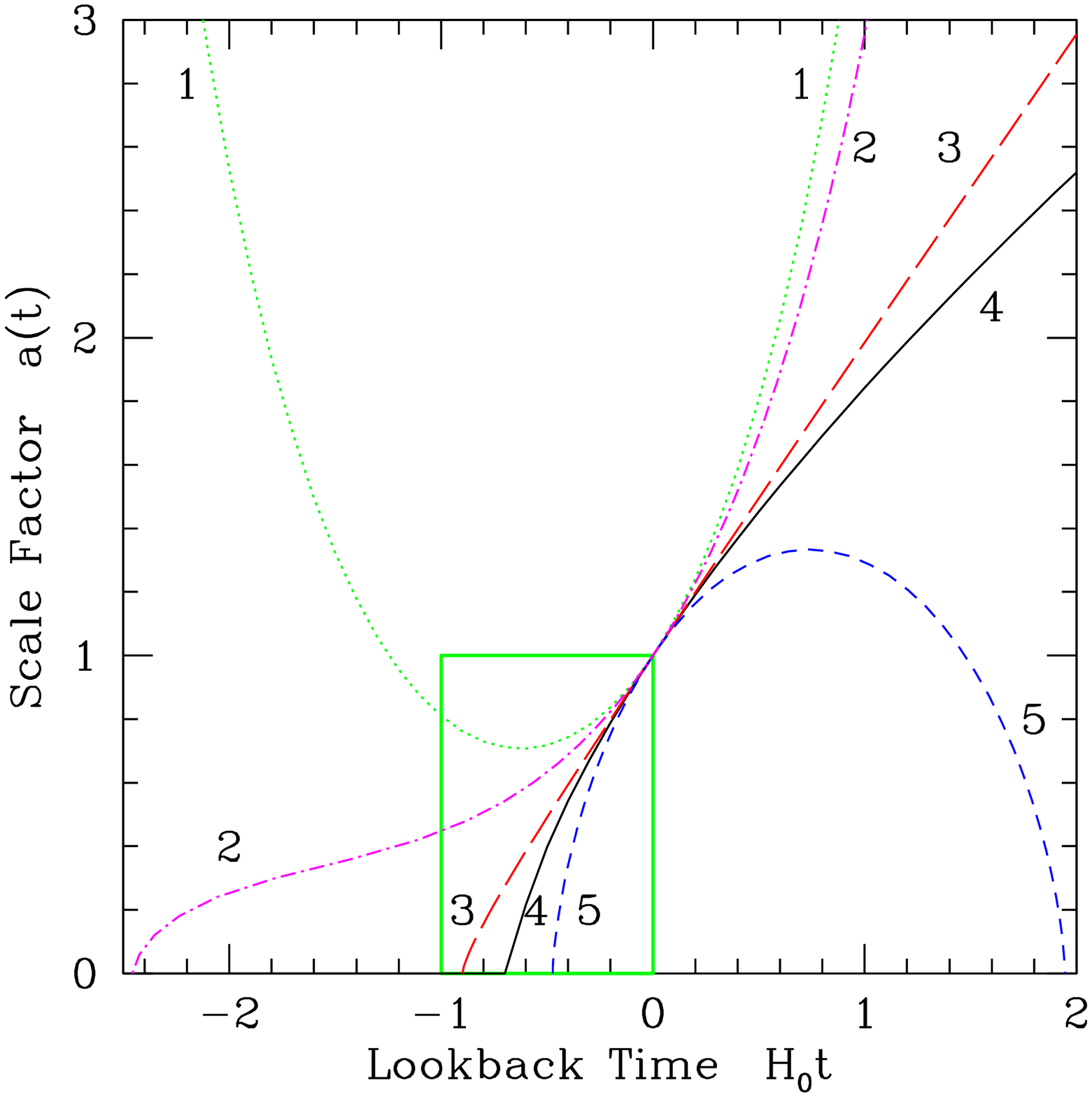,height=3.in} 
\psfig{file=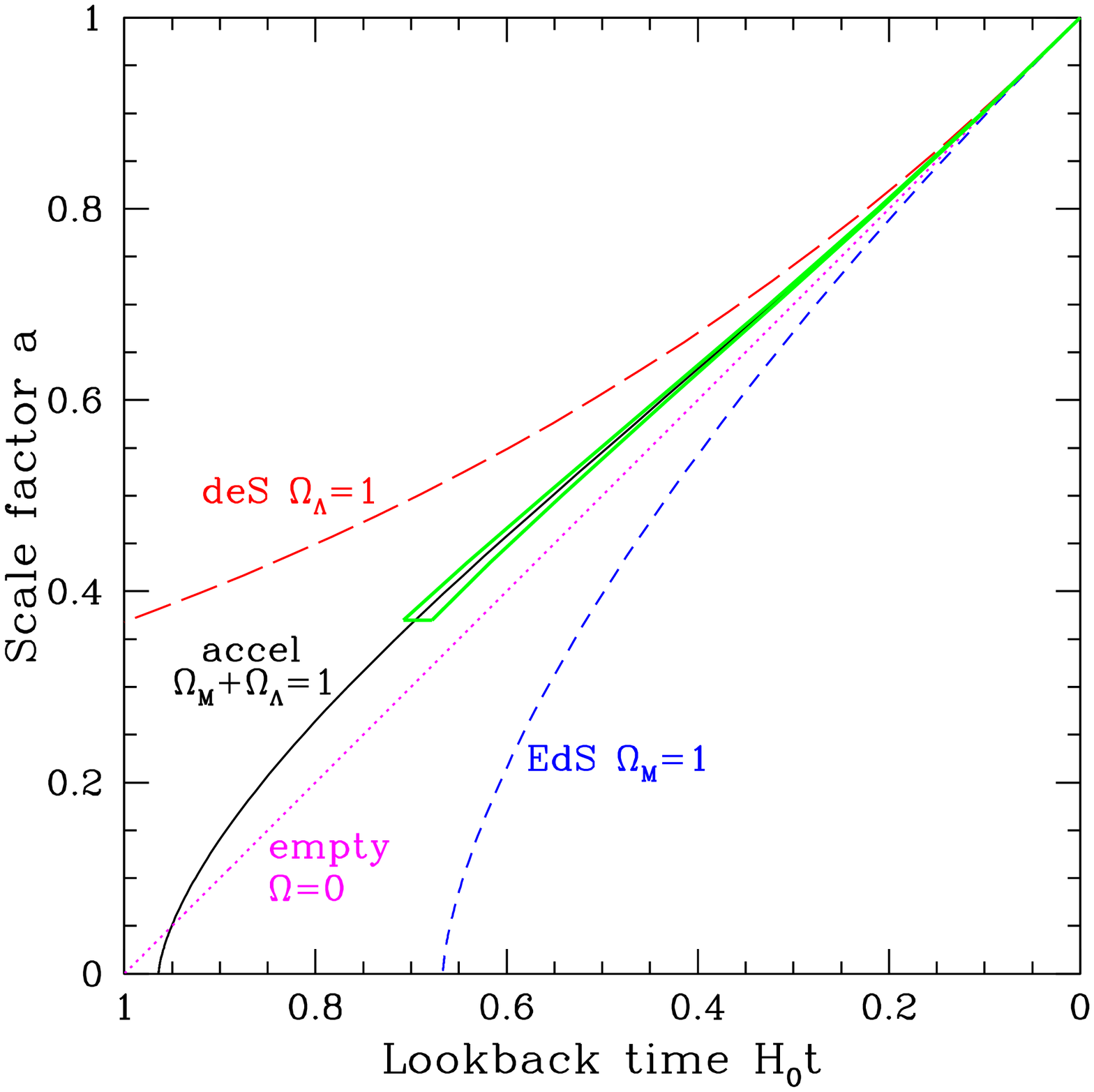,height=3.in} 
\psfig{file=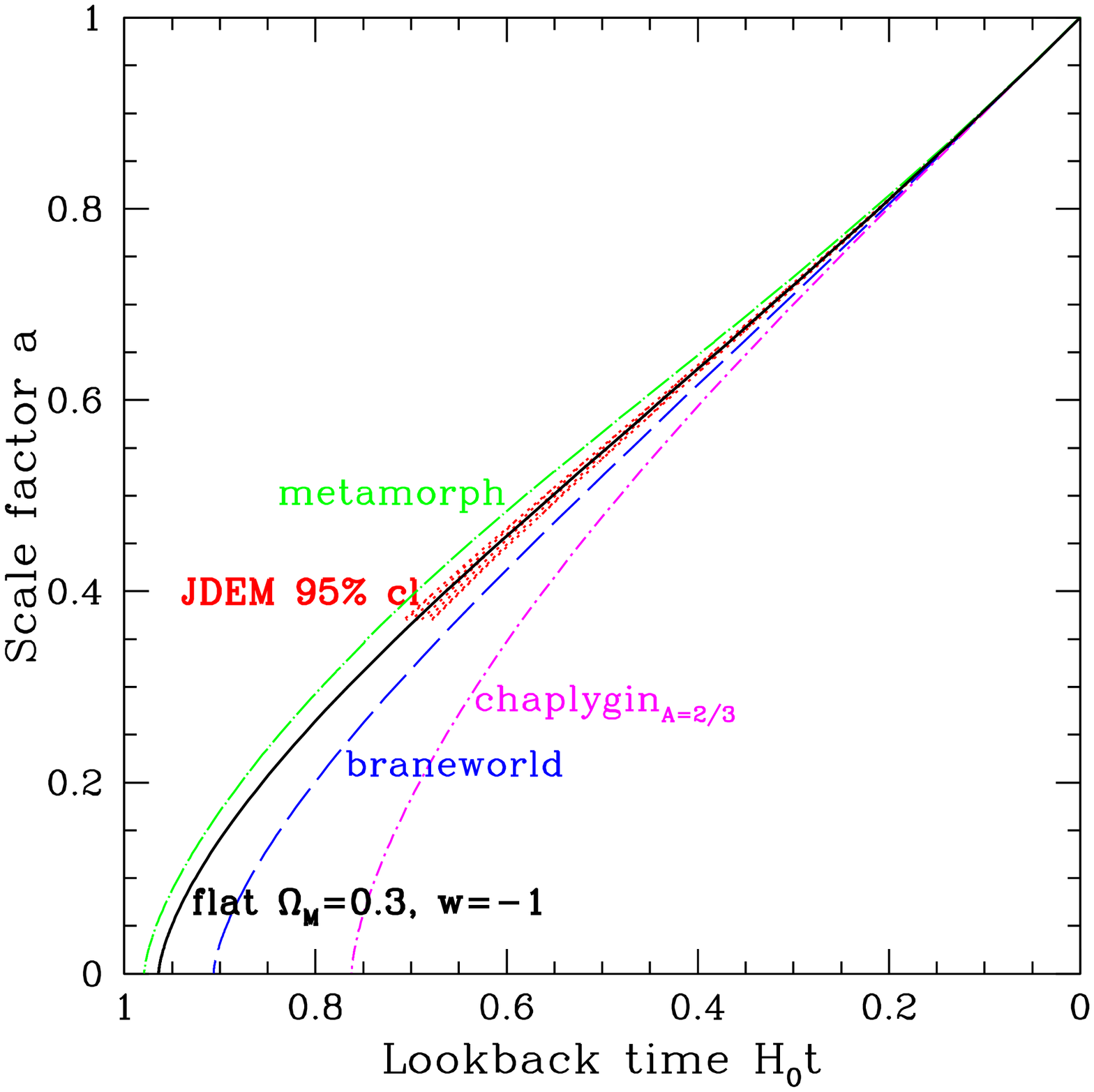,height=3.in} 
\caption{Mapping the expansion history has been a major theme of 
cosmology, revealing the constituents and nature of our universe. 
From early models with diverse properties (top left panel), scientists 
narrowed 
in to a region shown by the green box corresponding to Big Bang models 
with a long matter dominated epoch.  By the 1980s cosmologists 
believed the universe corresponded to an Einstein-de Sitter universe 
with critical matter density, or perhaps some model more toward the 
empty universe curve (top right panel).  The remarkable discovery in 1998 
of accelerated expansion showed that the correct model must be similar 
to the solid black curve, partaking of characteristics of the de Sitter 
universe with a cosmological constant $\Lambda$.  The challenge now is 
to further improve our cosmic mapping ability to zoom in on the narrow 
green triangular region, revealing the nature of the new physics behind 
acceleration (bottom panel).  (See \citet{Linprl} for the specific models.) 
}
\label{fig:history} 
\end{center} 
\end{figure}

From the 1930s to 1980s, astronomical observations of the behavior 
of sources at different redshifts, and the increasing corroboration of 
redshift as directly related to distance, confirmed the 
picture of an expanding universe.  (For a collection of some important 
early papers, see \citet{BernsteinJ}.)  Figure~\ref{fig:history} shows the 
evolution of our capabilities to map the cosmic expansion and the 
subsequent understanding achieved.  Comparison of theoretical calculations 
and observations with respect to primordial nucleosynthesis and the cosmic 
microwave background (CMB) radiation clarified the model as one arising from 
a hot, dense, near singular state given the name Big Bang.  However, 
estimates of the matter density (and even more so other component 
contributions, save for radiation measured directly in the CMB) remained 
somewhat vague, and in fact little different from Friedmann's 1922 work.  
Data could not make a definitive statement as to whether the matter and 
energy density amounted to less than, equal to, or possibly greater than 
the critical density needed for a spatially flat, asymptotically static 
expansion. 

This all changed dramatically in 1998 with the discovery of acceleration 
of the cosmic expansion, by two independent groups mapping the 
distance-redshift relation of Type Ia supernovae \citep{riess98,perl99}. 
The expansion history behavior jumped from being somewhere between a 
critical, matter dominated universe and an open, spatial curvature 
dominated universe (approaching an empty universe akin to the Minkowski 
space of merely special relativity), and instead shifted toward one with 
similarities to de Sitter space governed by a cosmological constant. 
What was so revolutionary was not that such behavior merely changes the 
quantitative aspects of the expansion, illustrated in 
Figure~\ref{fig:history}, but differs qualitatively and fundamentally, 
breaking the relation between geometry and destiny.

\subsection{Geometry and destiny} \label{sec:geom} 

For a non-accelerating universe described on large scales by a smooth, 
homogeneous and isotropic, i.e.\ Friedmann-Robertson-Walker model, 
geometry and destiny are inextricably linked.  Whether the spatial 
curvature is positive, zero (flat), or negative correlates one-to-one with 
whether the expansion is closed (bounded), critical (asymptotically static), 
or open (unbounded).  However, the Einstein field equations governing the 
cosmic expansion involve not only the amount of energy density but the full 
energy-momentum contribution.  So there is a loophole to escape destiny. 

The Friedmann 
equations (the Einstein equations in a homogeneous, isotropic model) can 
be written as 
\beq 
\dot a^2+k-8\pi G\rho a^2/3\equiv \dot a^2+V_{\rm eff}(a)=0, 
\eeq 
where $a$ is the expansion scale factor, $k$ is the spatial curvature, 
and $\rho$ is the total energy density.  (We consider the effective 
potential $V_{\rm eff}$ below.)  
From the continuity (or conservation of energy-momentum) equation, the 
total energy density $\rho$ will evolve with expansion as $\rho\sim 
a^{-3(1+w)}$, where $w=p/\rho$ is the pressure to energy density, or 
equation of state, ratio (easily generalized for a time varying ratio). 
So $\rho a^2\sim a^{-(1+3w)}$.  

If $w>-1/3$ then the sum of the $k$ and $\rho a^2$ terms 
ranges from $-\infty$ to $k$ for all values of $a$, i.e.\ the entire 
expansion history.  Thus, whether $\dot a^2$ ever reaches zero 
and hence whether there is a maximum value of $a$ or expansion 
continues without halt is wholly governed by the value of the spatial 
curvature $k$: is it positive or negative.  Geometry controls density. 

However, if $w<-1/3$ then the $\rho a^2$ term grows with $a$ and 
eventually dominates as the expansion factor grows large.  This strongly 
negative contribution can overcome a positive $k$, so this breaks the 
link between geometry and destiny.  

One can get a visual appreciation for this by considering the $k$ and 
$\rho a^2$ terms together as making up an effective potential energy 
(see, e.g., \citet{araa}), such as standardly used in physics analysis 
to determine whether a system with a certain kinetic energy is bound or 
not.  Here, the analog of kinetic energy is $\dot a^2$, and we want to 
know whether the minimum kinetic energy, $\dot a^2=0$, hits the potential 
energy curve for any value of $a$, indicating a maximum expansion factor. 

\begin{figure}[!htb]
\begin{center}
\psfig{file=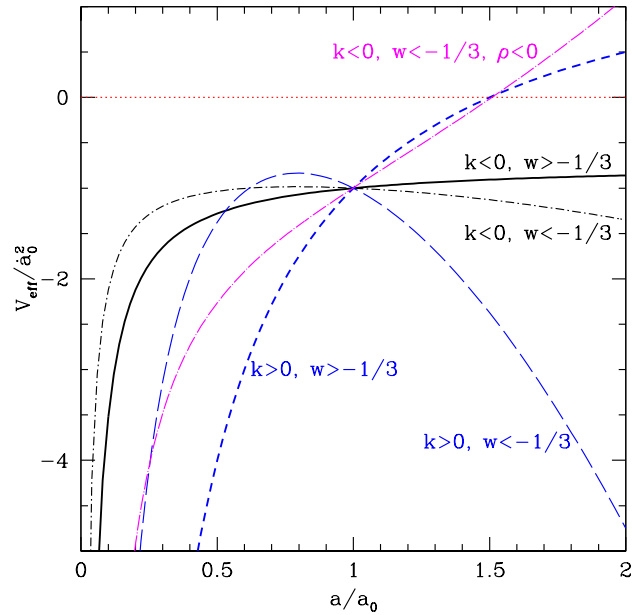,width=3.4in}
\caption{Effective potentials for universes classified according to 
their geometry (sign of $k$) and whether the cosmic expansion accelerates 
($w<-1/3$) or not.  Models crossing the dotted zero line do not expand 
forever.  For $w>-1/3$ the geometry determines the destiny: 
whether models with physical kinetic energy, hence $V_{\rm eff}\le0$, 
can achieve $a\to\infty$, i.e.\ expand forever.  By contrast, for 
accelerating universes the destiny is eternal expansion regardless of 
geometry -- unless the energy density $\rho$ can go negative. 
}
\label{fig:veff}
\end{center}
\end{figure}

Figure~\ref{fig:veff} for the effective potential $V_{\rm eff}$ vs.\ $a$ 
then illustrates the conditions discussed above.  If $w>-1/3$ then the 
effective potential 
ranges from $-\infty$ to $k$ for all values of $a$, i.e.\ the entire 
expansion history.  Thus, whether the effective potential crosses the 
minimum kinetic energy value of zero 
and hence whether there is a maximum value of $a$ or expansion 
continues without halt is wholly governed by the value of the spatial 
curvature $k$: geometry controls density.

However, if $w<-1/3$ then the second term of the effective potential 
eventually dominates as the expansion factor grows large.  This strongly 
negative contribution can overcome a positive $k$, so one can in fact have 
an eternal, positive curvature universe.  Conversely, if the energy density 
itself goes negative (e.g.\ due to a negative cosmological constant) then 
a finite, negative curvature universe is possible.  To determine the 
crucial quantity, the value of $w$, we need to take into account the 
entire energy-momentum not just the energy density.  
From the other Friedmann equation, 
\beq 
\frac{\ddot a}{a}=-\frac{4\pi G}{3}(\rho+3p)=-\frac{4\pi G}{3}\rho\,(1+3w), 
\eeq 
we see that the condition $w<-1/3$ is precisely the condition for 
accelerated expansion, $\ddot a>0$.  When one of the components of the 
universe has 
sufficiently negative pressure and contributes substantial energy density, 
such so-called dark energy can cause the total equation of state to drop 
below $-1/3$ and cause acceleration.  Destiny then becomes unhinged from 
geometry, and we must understand the nature of dark energy to predict the 
fate of the universe.  We discuss this further in \S\ref{sec:fate}.

\subsection{Acceleration} 

\begin{figure}[!htb]
\begin{center}
\psfig{file=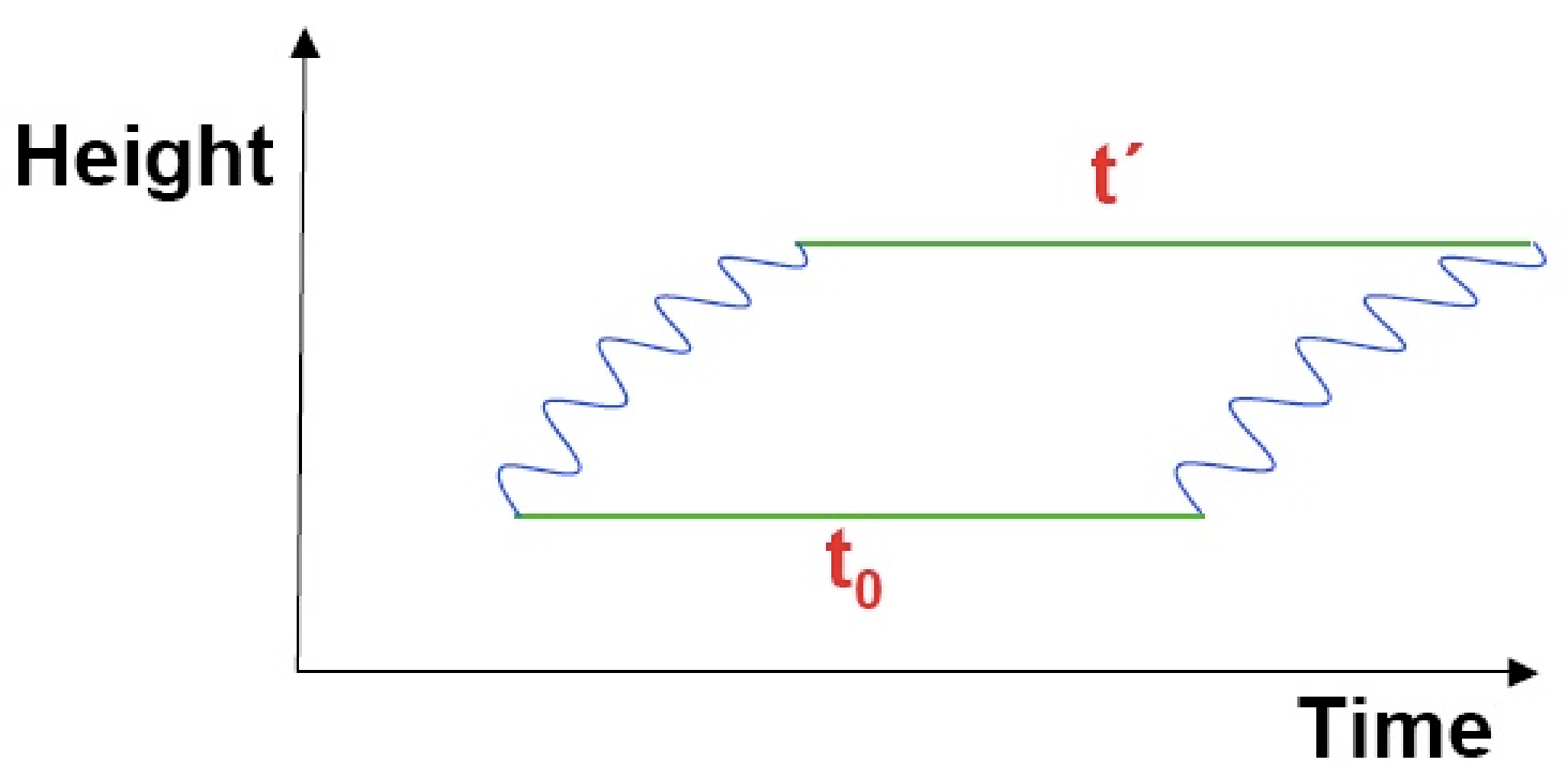,width=3.4in}
\caption{A photon experiences a frequency shift from acceleration, 
gravity, or spacetime curvature.  Here the spacetime diagram illustrates 
a Pound-Rebka experiment of a photon propagating from one atom to another 
at height $h$ above it.  A gravitational field leads to a gravitational 
redshift $z=gh/c^2$, while equivalently a uniform acceleration $g$ over a 
time $t$ leads to a velocity $v=gt$ and a Doppler shift $z=v/c=gt/c=gh/c^2$.  
Since a frequency change is equivalent to a change of time intervals, 
the horizontal, parallel line segments representing the emitted photon 
pulse period $t_0$ and received period $t'$ cannot be equal.  This means 
the photon propagation lines (wavy diagonal lines), which should each be at 
$45^\circ$ in a flat spacetime diagram, must not in fact be parallel.  Thus 
acceleration, gravitation, and spacetime curvature are equivalent. 
}
\label{fig:equiv}
\end{center}
\end{figure}

Einstein's equivalence principle states that acceleration is curvature 
is gravity.  While space may (or may not) be flat, {\it spacetime\/} 
curvature is nonzero in an expanding universe.  Rather than viewing gravity 
as a force acting at a distance, we can view it as the curvature of 
spacetime and particles follow paths of least action (geodesics) in this 
curved spacetime.  An intuitive way of seeing these deep connections is 
given in Figure~\ref{fig:equiv}. 

Thus, mapping the expansion history and its acceleration is equivalent to 
mapping spacetime or to exploring the gravitational nature of the universe.

\subsection{Revolution in physics} 

Is the discovery and further exploration of cosmic acceleration truly 
revolutionary?  Let us ask what is our current understanding of the 
universe through the Standard Model: baryons, photons, neutrinos, etc.\ 
make up roughly 4\% of the energy density of the universe.  Understanding 
4\% is like one letter out of the alphabet, e.g.\ reading 

\begin{quote} 
\fon{The nature} o\fon{f the physics} o\fon{f the accelerating universe 
is the premier mystery f}o\fon{r this generati}o\fon{n} o\fon{f physics.  
The challenge} o\fon{f a new, d}o\fon{minant, str}o\fon{ngly negative 
pressure c}o\fon{mp}o\fon{nent} o\fon{f the universe is cl}o\fon{sely 
tied with the f}o\fon{undati}o\fon{ns} o\fon{f quantum the}o\fon{ry and 
the nature} o\fon{f the vacuum; it is central t}o \fon{the the}o\fon{ry} 
o\fon{f gravitati}o\fon{n and the nature} o\fon{f spacetime, p}o\fon{ssibly 
even the number} o\fon{f dimensi}o\fon{ns.  The behavi}o\fon{r} 
o\fon{f dark energy g}o\fon{verns the fate} o\fon{f the universe.}  
%Understanding c}o{\fon smic acceleration undeniably extends the fr}o{\fon 
%ntiers} o{fon f physics and will rewrite the textb}oo{\fon ks}. 
\end{quote} 

We may hope to soon detect and eventually characterize dark matter.  
What will be our understanding once we have succeeded with this 
characterization and employed it to extend the Standard Model to a new 
theory of high energy physics such as supersymmetry?  Then we will 
comprehend 25\% of the energy density of the universe, like all the 
vowels (with y) of the alphabet: 

\begin{quote}
\fon{Th}e \fon{n}a\fon{t}u\fon{r}e o\fon{f th}e \fon{ph}y\fon{s}i\fon{cs} 
o\fon{f th}e a\fon{cc}e\fon{l}e\fon{r}a\fon{t}i\fon{ng} 
u\fon{n}i\fon{v}e\fon{rs}e i\fon{s th}e \fon{pr}e\fon{m}ie\fon{r 
m}y\fon{st}e\fon{r}y \fon{f}o\fon{r th}i\fon{s 
g}e\fon{n}e\fon{r}a\fon{t}io\fon{n} o\fon{f ph}y\fon{s}i\fon{cs.  Th}e 
\fon{ch}a\fon{ll}e\fon{ng}e o\fon{f} a \fon{n}e\fon{w, 
d}o\fon{m}i\fon{n}a\fon{nt, str}o\fon{ngl}y \fon{n}e\fon{g}a\fon{t}i\fon{v}e 
\fon{pr}e\fon{ss}u\fon{r}e \fon{c}o\fon{mp}o\fon{n}e\fon{nt} o\fon{f th}e 
u\fon{n}i\fon{v}e\fon{rs}e i\fon{s cl}o\fon{s}e\fon{l}y \fon{t}ie\fon{d 
w}i\fon{th th}e \fon{f}ou\fon{nd}a\fon{t}io\fon{ns} o\fon{f 
q}ua\fon{nt}u\fon{m th}eo\fon{r}y a\fon{nd th}e \fon{n}a\fon{t}u\fon{r}e 
o\fon{f th}e \fon{v}a\fon{c}uu\fon{m;} i\fon{t} i\fon{s c}e\fon{ntr}a\fon{l 
t}o \fon{th}e \fon{th}eo\fon{r}y o\fon{f gr}a\fon{v}i\fon{t}a\fon{t}io\fon{n} 
a\fon{nd th}e \fon{n}a\fon{t}u\fon{r}e o\fon{f sp}a\fon{c}e\fon{t}i\fon{m}e\fon{, p}o\fon{ss}i\fon{bl}y e\fon{v}e\fon{n th}e \fon{n}u\fon{mb}e\fon{r} 
o\fon{f d}i\fon{m}e\fon{ns}io\fon{ns.  Th}e \fon{b}e\fon{h}a\fon{v}io\fon{r} 
o\fon{f d}a\fon{rk} e\fon{n}e\fon{rg}y \fon{g}o\fon{v}e\fon{rns th}e 
\fon{f}a\fon{t}e o\fon{f th}e u\fon{n}i\fon{v}e\fon{rs}e\fon{.}  
\end{quote}

It is apparent that for true understanding we will need to know the nature of 
dark energy.  Only then will we have a complete and comprehensible picture: 

\begin{quote} 
The nature of the physics of the accelerating universe is the premier 
mystery for this generation of physics.  The challenge of a new, dominant, 
strongly negative pressure component of the universe is closely tied with 
the foundations of quantum theory and the nature of the vacuum; it is 
central to the theory of gravitation and the nature of spacetime, possibly 
even the number of dimensions.  The behavior of dark energy governs the 
fate of the universe.  
\end{quote} 

Understanding cosmic acceleration undeniably will extend the frontiers 
of physics and rewrite the textbooks.  It is not excessive to suggest 
that we are faced with a revolution in our understanding of nature as 
profound as the mystery of blackbody radiation a century ago.  Blackbody 
radiation taught us the existence of photons and the structure of atoms; 
dark energy may lead us to the existence of quantum gravity and the 
structure of the vacuum.

\section{Effects of accelerated expansion} \label{sec:eff} 

\subsection{Acceleration directly?} 

Given the importance of cosmic acceleration, is there some way to detect 
it in itself, rather than through the curves of the expansion history? 
Recall Figure~\ref{fig:equiv}.  There we motivated the Equivalence Principle 
by showing how acceleration equals gravity.  We can detect the 
acceleration through the cosmological version of the gravitational 
redshift.  Photons (or any signal) will have their frequencies shifted 
by the acceleration just as they would by a gravitational field.  Of course 
the expansion of space in itself redshifts photons, but that is analogous 
to a velocity or Doppler shift; acceleration will add a second time 
derivative of the photon frequency, showing up a drift in the redshift $z$. 

The redshift drift was first discussed by \citet{sandage62} and given in 
general form by \citet{mcvittie62}.  Analysis of its use as a cosmological 
probe, including observational challenges and systematics, appeared in 
\citet{lindrift,fpoc}.  The result is that 
\beq 
\dot z=H_0(1+z)-H(z)=a^{-1}\,[H_0-\dot a], 
\eeq 
where $H=\dot a/a$ is the Hubble parameter and $H_0$ is the present 
expansion rate, the Hubble constant.  Since the scale factor $a=1/(1+z)$ 
we clearly see that $\dot z$ vanishes only in universes with 
$\dot a={\rm constant}$.  That is, redshift drift is a direct 
signature of acceleration (or deceleration).  However, since the Hubble 
time $H^{-1}$ is more than 10 billion years, 
the redshift drifts at only 1 part in $10^{10}$ per year, beyond present 
technology to measure.  Even with a 20 year observational program with 
stability achieved at the one part in a billion level, the cosmological 
leverage of such a measurement is unimpressive.  Moreover, as 
\citet{lindrift,fpoc,UzanDrift} 
pointed out, just as peculiar velocities interfere with accurate redshift 
measurements, so would peculiar accelerations degrade redshift drift.  
These can take the form of endpoint effects, i.e.\ jitter in the observer 
or source motion due to realistically inhomogeneous gravitational fields 
from mass flows, or propagation effects such as a stochastic 
gravitational wave background with energy density as small as $10^{-17}$ 
of critical density would generate significant noise. 

As for dynamical effects of dark energy within the solar system or 
astrophysical systems, the energy density is simply too low.  All the 
dark energy within the entire solar system constitutes the energy equivalent 
of three hours of sunlight at 1 AU, ruling out any direct effect on orbits. 
For lensing by black holes, say, the relative contribution to the deflection 
of light by a cosmological constant $\Lambda$ is $\Lambda r^2/(m/r)$.  
Detectable lensing requires a gravitational potential $\Phi\sim m/r\gs  
10^{-6}$, and $\Lambda\sim H^2$ so it would require a black hole with 
mass greater than $10^{13}\,M_\odot$ before dark energy would contribute 
even 1\% effect -- and at that point the static approximation used here 
breaks down.  Thus, cosmological expansion remains the practical path to 
mapping acceleration.

\subsection{Kinematics} 

It is attractive to consider how much information one can extract on 
the expansion history from minimal assumptions.  If one uses only the 
geometry of spacetime as an input, i.e.\ the Robertson-Walker metric, 
then one can learn a remarkable amount.  For example, the expansion of 
space and redshift of photons, hence the decrease in temperature and 
density as the universe expands, are directly derived from examination 
of the metric. 
Such properties that do not rely on supplementing the geometry with 
equations of motion are referred to as kinematics, while those that 
depend on the field equations, i.e.\ the specific theory of gravity, 
fall under dynamics. 

For example, the relation between conformal distance $\eta$ and the 
expansion factor $a$ is kinematic, 
\beq 
\eta(a_\star)=\int_{a_\star}^1 \frac{da}{a^2H}, 
\eeq 
but if $H$ is defined in terms of $\eta$, i.e.\ $H=\dot a/a=a^{-2}da/d\eta$, 
then this becomes a tautology.  To actually evaluate the distance-redshift 
relation requires a model for $H$, generally supplied by the equations of 
motion, that is the Friedmann equations in terms of the densities and 
pressures.  Before returning to this, let us consider kinematic signatures 
of acceleration. 

\begin{figure}[!htb]
\begin{center}
\psfig{file=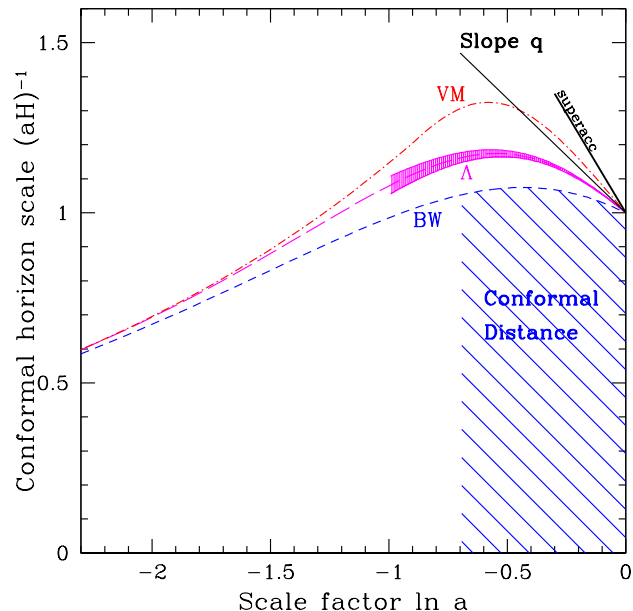,width=3.4in}
\caption{This conformal history diagram presents a unified picture of 
important properties of the cosmic expansion.  Curves represent the 
expansion history of different cosmological models (here the cosmological 
constant $\Lambda$, an extra dimension braneworld scenario, and a vacuum 
metamorphosis or quantum phase transition).  The value of a point on a 
curve measures the conformal horizon, basically the size of the universe, 
as a function of the cosmic expansion factor. The slope of the curve gives 
the deceleration parameter $q$; if it is positive then the universe is 
decelerating but if the slope is negative the universe is accelerating.  
The area under a curve gives the distance measured by observers from the 
present ($a=1$ or $\ln a=0$) to some time in the past.  The shading around 
the $\Lambda$ curve shows how well next generation experiments should probe 
the cosmic expansion. 
}
\label{fig:ahinv}
\end{center}
\end{figure}

In the diagrams of Figure~\ref{fig:history}, acceleration would show up 
as convexity of the curves, i.e.\ the second derivative becomes positive. 
A clearer and more direct way of seeing acceleration is by transforming 
the variables plotted to the conformal horizon scale as a function of 
expansion factor; see Figure~\ref{fig:ahinv}.  A size of the visible 
universe can be defined in terms 
of the Hubble length, $H^{-1}$, where the expansion, or e-folding, rate 
$H=\dot a/a$.  Since all lengths expand with the scale factor $a$, we 
can transform into conformal, or comoving, coordinates by dividing by 
$a$, thus defining the conformal horizon scale $(aH)^{-1}$. 

Comoving wavelengths would appear as horizontal lines in this plot and 
so wavelengths enter the horizon, i.e.\ fall below the horizon history 
curve, at early times as the universe expands.  For decelerating expansion, 
this is the whole story, with more wavemodes being revealed as the 
expansion continues.  However, for accelerating universes, the slope of 
the horizon history curve goes negative and modes can leave the horizon. 
This condition $d(aH)^{-1}/d\ln a\sim -\ddot a<0$ and its consequences 
are precisely 
the principle behind inflation: an accelerating epoch in the early universe. 

This diagram demonstrates visually the tight connection between the 
expansion history (the value along the curve), acceleration (the slope 
of the curve), and distances of objects in the universe (the area 
under the curve).  Thus distances provide a clear and direct method 
for mapping the expansion history, and will be treated at length in 
\S\ref{sec:dist}. 

However, as stated above, to actually evaluate the curve for an expansion 
history one needs a model for the expansion.  One could adopt an ad hoc 
model $a(t)$ or $H(z)$ or, parameterizing the acceleration directly, $q(z)$ 
where $q=-a\ddot a/\dot a^2$ is called the deceleration parameter, or 
even the third derivative $j=a^2\dddot a/\dot a^3$ called the jerk.  One 
runs the danger of substituting the physics of the Einstein equations with 
some other, implicit dynamics since adopting a form for $q(z)$ is equivalent 
to some ad hoc equation of motion.  Explicitly, if one defines $H^2=f(\rho)$, 
some function of the total density, say, then the continuity equation leads 
to $q=-1+(3/2)(1+w)\,d\ln f/d\ln\rho$, so choosing some functional form 
$q(z)$, or $j(z)$, {\it is\/} choosing an equation of 
motion.  That is, one has not truly achieved kinematic constraints, 
only substituted some other unspecified physics for general relativity.  

One method for getting around this is by putting 
no physics into the form by taking a Taylor expansion, e.g.\ $q(z)=q_0+q_1z$ 
\citep{Blandfordkin,Aminkin}.  
This has limited validity, that is it can only be applied to mapping the 
expansion at very low redshifts $z\ll1$ and it is not clear what has actually 
been gained.  Another method is to allow the data to determine the form, 
and the physics, through principal component analysis as in 
\citet{ShapTur,Aminkin}.  This has a broader range of physical validity, but 
has substantial sensitivity to noise in the data, since it seeks to extract 
information on a third derivative in the case of jerk.  

\subsection{Dynamics} \label{sec:dyn} 

Einstein's equations provide the dynamics in terms of the energy-momentum 
components in the universe.  Note that alterations to the form of the 
gravitational 
action also define the dynamics.  Unless otherwise specified we consider 
Friedmann-Robertson-Walker cosmologies.  In this case, the simplest 
ingredients are the energy density and pressure of each component, with 
the components assumed to be noninteracting.  This can be phrased 
alternately in terms of the present dimensionless energy density 
$\Omega_w=8\pi G\rho_w(z=0)/(3H_0^2)$ and the pressure to density, or 
equation of state, ratio $w=p_w/\rho_w$. 

\begin{figure}[!htb]
\begin{center}
\psfig{file=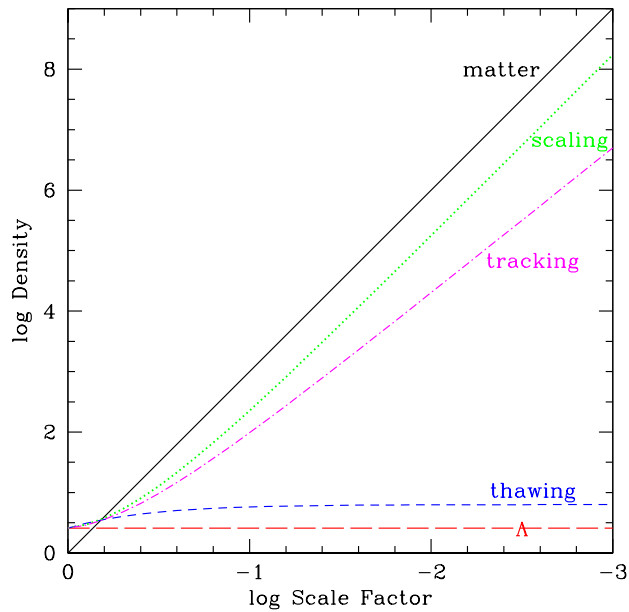,width=3.4in}
\caption{In the recent expansion history, matter and dark energy have 
contributed similar amounts of energy density (shown here normalized 
to the present matter density) but this is apparently 
coincidental.  Among quintessence models, the energy density of the 
cosmological constant or a thawing field differs substantially at early 
times from freezing fields such as trackers or scalers, one example of 
classification of dark energy models. 
}
\label{fig:rhot}
\end{center}
\end{figure}

We have seen in \S\ref{sec:geom} that the equation of state is central 
to the relation between geometry and destiny, and it plays a key role 
in the dynamics of the expansion history as well.  In Figure~\ref{fig:rhot} 
we see the very different behaviors for the dark energy density for 
different classes of equations of state.  The cosmological constant has 
an unchanging energy density, and so it defines a unique scale, tiny 
in comparison to the Planck energy, and a particular time in cosmic 
history when it is comparable to the matter density.  These fine tunings 
give rise to the cosmological constant problem 
\citep{Wbg89,CarrollLiv,BoussoGRG}. 
Dynamical scalar fields, called quintessence, change their energy 
density and this may or may not alleviate the large discrepancy with the 
Planck scale.  One of the two main classes of such fields \citep{CaldLin}, 
thawing fields that evolve away from early cosmological constant behavior, 
has energy density that changes little.  The other class, freezing 
fields that evolve toward late time cosmological constant behavior, 
can have much greater energy density at early times.  For example, tracking 
fields may contribute an appreciable fraction of the energy density 
over an extended period in the past and may approach the Planck density 
at early times, while scaling fields mimic the behavior of 
the dominant energy density component, contributing so much to the dynamics 
that they can be strongly constrained.

Although there is a diversity of dynamical behaviors \citep{LinGRG}, these 
fall into distinct classes of the physics behind acceleration \citep{CaldLin}. 
We see that to reveal the origin of cosmic acceleration we will require 
precision mapping of the expansion history over the last e-fold of 
expansion, but we may also need to measure the expansion history at early 
times.  And to predict the future expansion history and the fate of the 
universe requires truly understanding the nature of the new physics -- 
for example knowing whether dark energy will eventually fade away, 
restoring the link between geometry and destiny.

\subsection{True acceleration?} 

Before proceeding further we can ask whether the Robertson-Walker model 
of a homogeneous, isotropic universe is indeed the proper framework for 
analyzing the expansion history.  Certainly the global dynamics of the 
expansion follows that of a Friedmann-Robertson-Walker model (FRW), as the 
successes of Big Bang nucleosynthesis, cosmic microwave background 
radiation measurements, source counts etc.\ show \citep{pdg,pdgpar}, 
but one could imagine smaller scale inhomogeneities affecting the light 
propagation by which we measure the expansion history.  This has long 
been known \citep{Sachs,KristianSachs,Gunn67} and for stochastic 
inhomogeneities 
shown to be unimportant in the slow motion, weak field limit \citep{jlw}. 

To grasp this intuitively, consider that the expansion rate of space is 
not a single number but a $3\times3$ matrix over the spatial coordinates. 
The analog of the Hubble parameter is (one third) the trace of this matrix, 
so inhomogeneities capable of altering the global expansion so as to 
mimic acceleration generically lead to changes in the other matrix 
components.  This induces shear or rotation of the same order as the 
change in expansion, leading to an appreciably anisotropic universe. 
Observations however limit shear and rotation of the expansion to be 
less than $5\times 10^{-5}$--$10^{-6}$ times the Hubble term \citep{rotation}. 

Thus, to create the illusion of acceleration one would have to carefully 
arrange the material contents of the universe, adjusting the density 
along the line of sight (spherical symmetry is not wholly necessary 
with current data quality).  Again, this has long been discussed 
\citet{FeynmanLight,ZeldovLight,DyerRoeder} and indeed changes the 
distance-redshift relation.  The simple model of \citet{Lin9801} poses the 
problem in its most basic terms, clearly demonstrating its meaning.  
It considers an inhomogeneous, matter only universe with a void 
($\alpha=0$ for the Dyer-Roeder smoothness parameter) somewhere along 
the line of sight, extending from $z_1$ to $z_1+\Delta z$, 
and finds that the distances to sources lying at higher redshift do not 
agree with the FRW relation.  Even for very high redshift sources where the 
cosmic volume is essentially wholly described by FRW there maintains an 
asymptotic 
fractional distance deviation of $\Delta z/(1+z_1)$ (if the void surrounds 
the observer then the deviation is of order $\Delta z^2$). 

\begin{figure}[!htb]
\begin{center}
\psfig{file=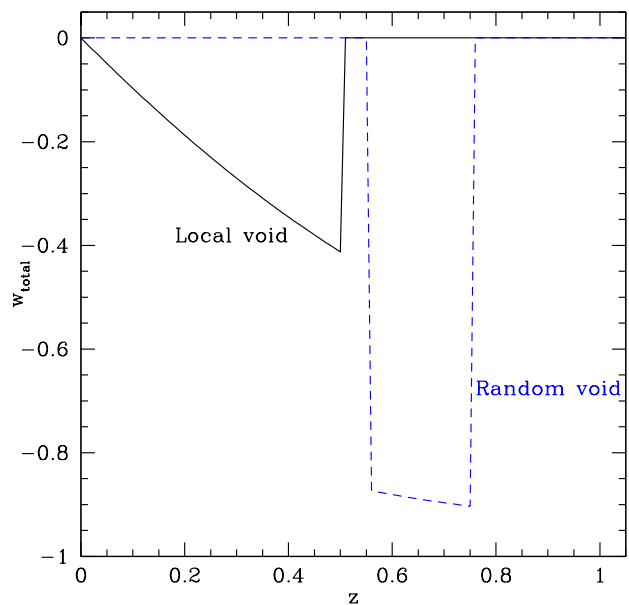,width=3.4in}
\caption{Huge voids can give the illusion of an effective negative 
equation of state and acceleration, but not the dynamical reality. 
The solid curve gives the effective total equation of state if we observe 
from the center of a void extending out to $z=0.5$, yet interpret data 
within a smooth FRW model.  The dashed curve corresponds to a void in 
a shell from $z=0.55-0.75$. 
}
\label{fig:voidw}
\end{center}
\end{figure}

Suppose we were to define $w(z)$ in terms of derivatives of distance with 
respect to redshift.  Then we would find for certain choices of void 
size and location that $w(z)<-1/3$; see Figure~\ref{fig:voidw}.  Apart from 
the fact that this would require enormous voids, it would be a mistake to 
interpret this as apparent acceleration within this very basic, matter 
only model.  The analysis is physically inconsistent because it treats the 
expansion, e.g.\ $H(z)$, in two different ways: as dynamics, for example in 
the friction term in the Raychaudhuri or beam equation, and as kinematics, 
through the correspondence to the differential of the distance, 
$dr/dz\sim H$.  We emphasize this point: in models with inhomogeneities, 
it is inconsistent to treat dynamics and kinematics the same.  
See \citet{IshibashiWald} for further discussion of the proper physical 
interpretation of acceleration.

While the previous approach can deliver a mirage of, but not physical, 
acceleration, one cannot even obtain such an illusion if one requires 
other basic aspects of FRW to hold.  From the Raychaudhuri, or beam, 
equation (see \citet{Sachs}) 
in FRW generalized to arbitrary components and smoothness \citep{Lin88ra}, 
one finds the following two conditions, 
\beq 
\om(z)=1+3w_{\rm tot}(z) \quad ;\quad \alpha(z)=\frac{1+w_{\rm tot}}{1+ 
3w_{\rm tot}}\,, 
\eeq 
must hold for a pure matter, inhomogeneous model to mimic a smooth model 
with matter plus an extra component with equation of state $w$ (so the total 
equation of state is $w_{\rm tot}$).  These follow from matching the 
Raychaudhuri equation term by term between the models, 
so that the distance-redshift relations will agree.  However, the 
requirement of positive energy density then immediately 
demonstrates that an effective $w_{\rm tot}<-1/3$ cannot be attained, i.e.\ 
acceleration is not possible.  Moreover, the matter in the inhomogeneous 
dust model cannot consistently obey the usual continuity equation.  

Thus, a pure dust model with inhomogeneities described by spatial under- 
and overdensities, i.e.\ $\alpha(z)$, cannot give distances matching an 
accelerating FRW universe distance-redshift relation, nor any FRW model 
without introducing nonstandard couplings in the matter sector. 

Finally, to obtain even the mirage of a perfectly isotropic 
distance-redshift relation requires 
the inhomogeneities to be arranged in a spherically symmetric manner 
around the observer, raising issues of our preferred location.  If the 
inhomogeneities are arranged only stochastically, i.e.\ do not have an 
infinite coherence 
length caused by special placement, then the effects along the line of 
sight will average out and the distance-redshift relation will reflect 
the true global expansion.  Thus, save for hand fashioning a universe to 
deceive, observed acceleration is real acceleration. 

\subsection{Fate of the universe} \label{sec:fate} 

Given accelerated expansion at present, we still cannot absolutely 
predict the future expansion and the fate of the universe.  If the 
acceleration from dark energy continues then the universe becomes a 
truly dark, cold, empty place.  The light horizon, within which we 
can receive signals, grows linearly with time, as always, but the 
particle horizon giving the (at some time) causally influenced region 
grows more quickly in an accelerating universe (exponentially in the 
cosmological constant case).  Thus, though formally our observational 
reach out into the universe increases, we see an ever tinier 
fraction of the causal universe.  (See \citet{TamHoriz} for more 
on horizons.) 

Also, although the light horizon expands, in a real sense the visible 
universe does ``close in'' around us not through objects going beyond 
the horizon but through their fading away to our sight.  For example, 
in a cosmological constant dominated universe the redshift 
of an object at constant comoving distance increases exponentially 
with time, so its received flux decreases exponentially and its surface 
brightness fades exponentially from the usual $(1+z)^4$ law.  
Conversely, the comoving distance to a fixed redshift decreases 
exponentially with time, so vanishingly few objects lie within the 
volume to any finite redshift and hence have non-infinitesimal 
flux and surface brightness.  (See \citet{KraStar,KraSch,araa} for 
some specific 
calculations.)  
Thus our view of the universe does not so much shrink as darken. (So 
dark energy is well named.)   

The existence of a horizon arising from acceleration, or Rindler 
horizon \citep{Rindler}, 
brings another set of physics puzzles, best known for the borderline case 
of $w=-1$.  In de Sitter space the horizon causes loss of unitarity and 
makes it problematic to define particle interaction probabilities through 
an S matrix \citep{Wbg89}.  On the other hand, the horizon may be 
instrumental in 
obviating the Big Rip fate.  For a universe with $w<-1$ the increasing 
(conformal) acceleration overcomes all other binding forces, ripping 
planets, atoms, etc.\ apart \citep{Bigrip}.  However, 
Unruh radiation, or thermal particle creation from the horizon, 
with temperature $T\sim g$ and hence energy density $\rho\sim T^4$ 
where $g=\ddot a/a$ is the conformal acceleration (so the rip condition 
is not $\dddot a>0$ but 
$(\ddot a/a)\,\dot{}>0$), should quickly overwhelm the dark energy 
(whose density increases as $g$, not $g^4$).  Thus the horizon should 
act to either decelerate the universe or bring the expansion to some 
non-superaccelerating state \citep{Lin04}. 

Other possibilities for the fate of the universe include collapse ($a\to0$) 
if the dark energy attains negative values of its potential, as in 
the linear potential model \citep{Linde86} (also see \S\ref{sec:doom}), 
collapse and bounce into new expansion as in the cyclic 
model \citep{ekpyrotic}, or eternal but decelerating expansion if the 
dark energy fades away. 

Thus we have seen throughout this section that to distinguish the origin 
of cosmic acceleration 
we may need not only to map accurately the {\it recent\/} expansion 
history, but distinguish models of dark energy through their {\it early\/} 
time behavior and understand their nature well enough to predict their 
{\it future\/} evolution. 
To truly understand our universe we must map the cosmic expansion from 
$a=0$ to $a=\infty$ -- and possibly back to $a=0$ again.

\section{Distance measures} \label{sec:dist} 

Distance measurements as a function of scale factor or redshift directly 
map the expansion history.  Here we consider several approaches to these 
measurements, starting with the geometric or mostly geometric methods 
that give the cleanest probes of the expansion.  Many other techniques 
involving distances exist, also containing noncosmological quantities. 
One can categorize probes into ones depending (almost) exclusively 
on geometry, ones requiring some knowledge of the mass of objects to 
separate out the distance dependence, and ones requiring not only knowledge 
of mass but the thermodynamic or hydrodynamical state of the material, 
i.e.\ mass+gas probes.  The discussion here focuses on the geometric 
distance techniques actually implemented to date, though we do briefly 
mention some other approaches. 

\subsection{General cosmological distance properties} \label{sec:distgen} 

\begin{figure}[!htb]
\begin{center}
\psfig{file=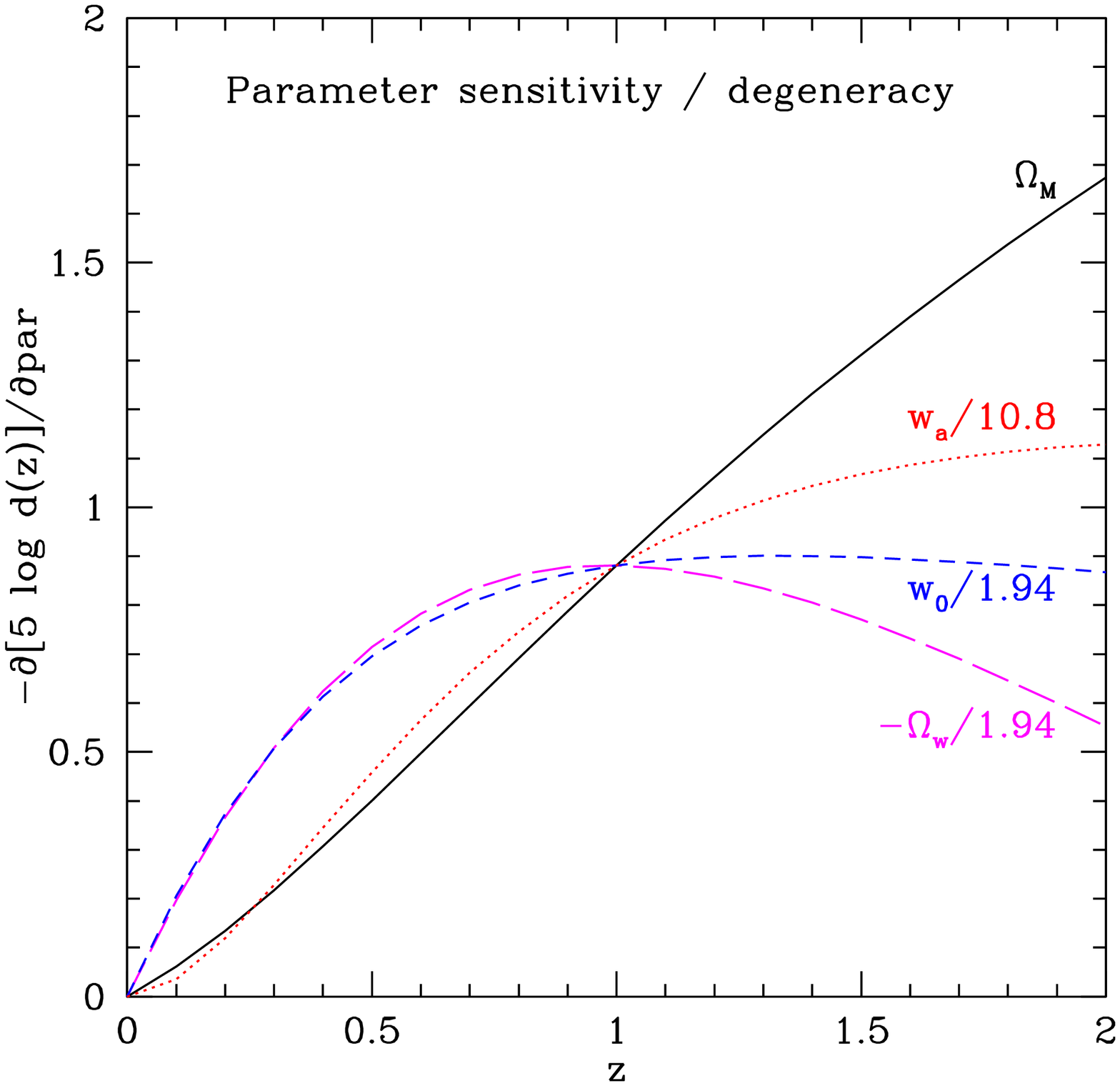,width=3.in}
\psfig{file=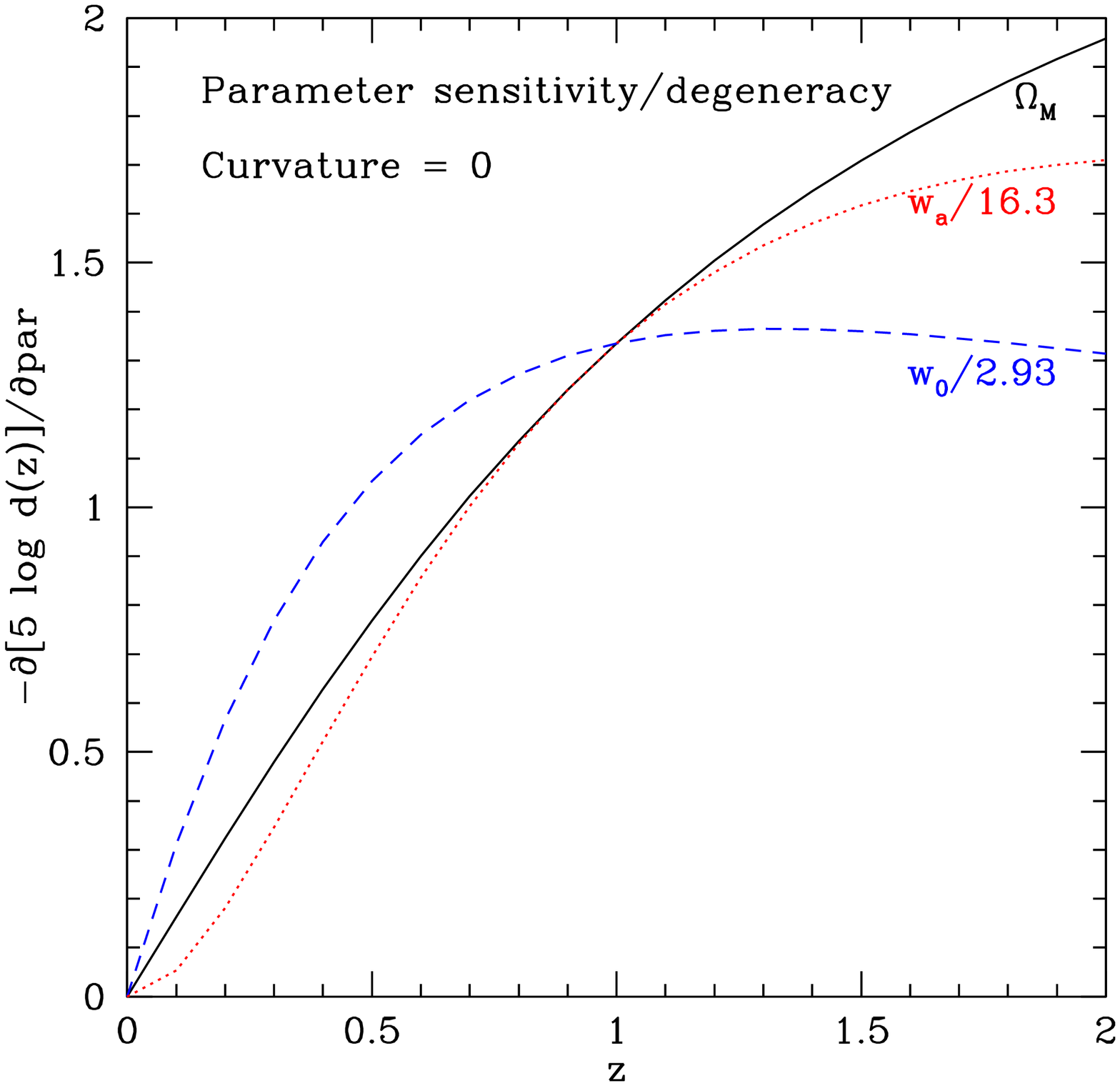,width=3.in}
\caption{Fisher sensitivities for the magnitude ($5\log d$) are plotted 
vs.\ redshift for the case leaving spatial curvature free (left panel) 
and fixing it to zero, i.e.\ a flat universe (right panel).  Since only 
the shapes of the curves matter for degeneracy, parameters are normalized 
to make this more evident.  Observations out to $z\ge1.5$ are required 
to break the degeneracies. 
}
\label{fig:sensd}
\end{center}
\end{figure}

\begin{figure}[!htb]
\begin{center}
\psfig{file=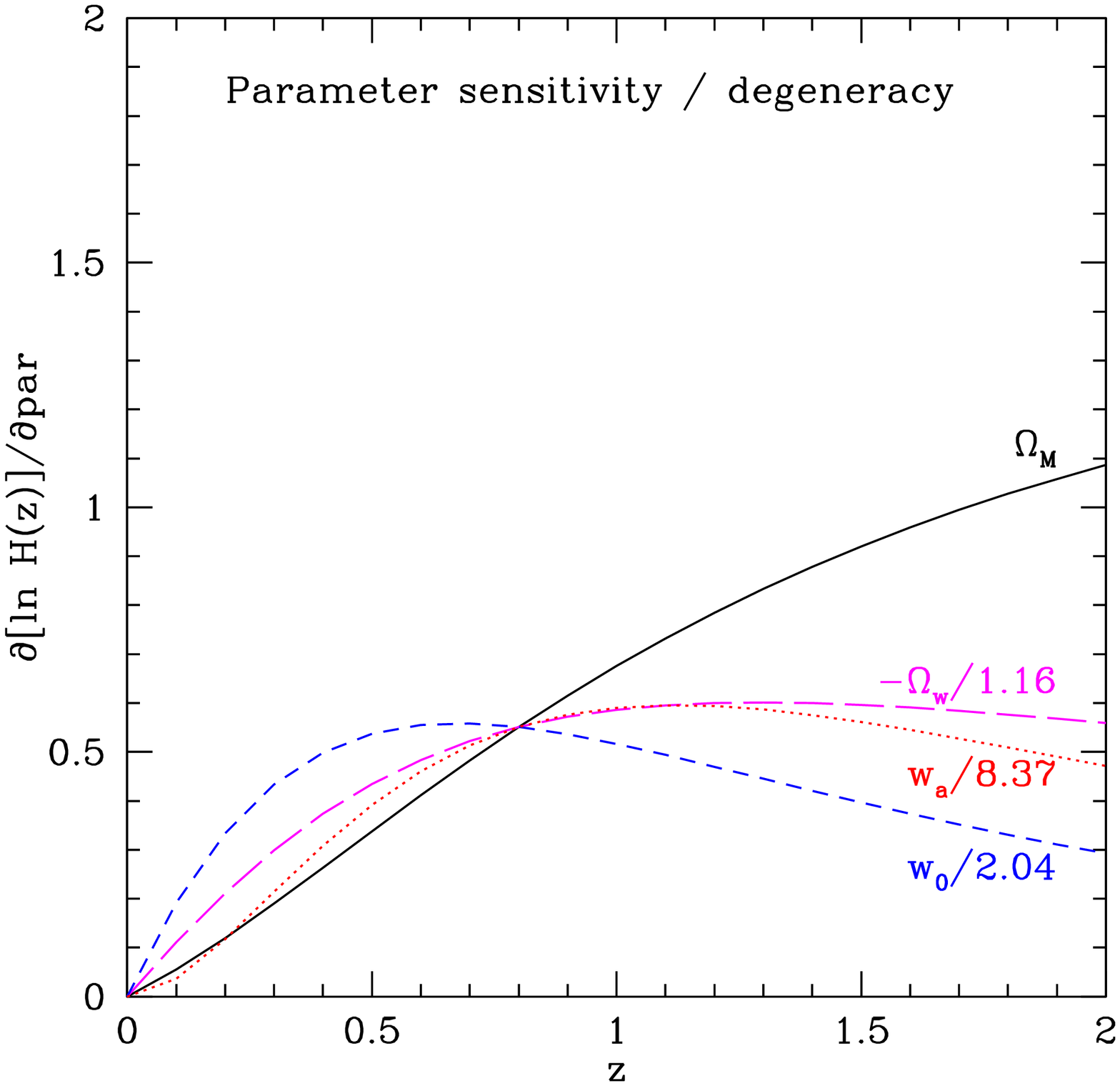,width=3.in}
\psfig{file=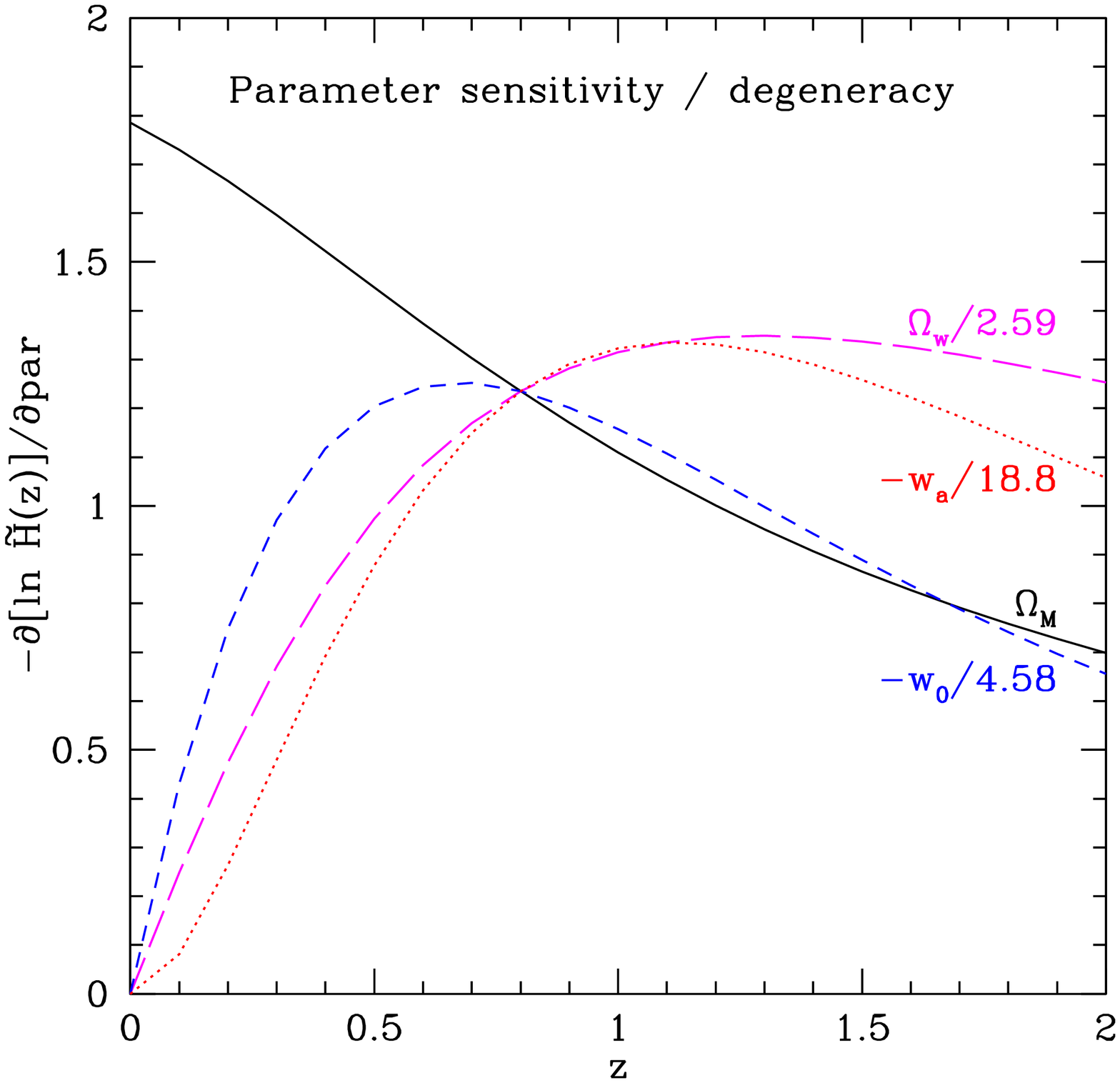,width=3.in}
\caption{Fisher sensitivities for the Hubble parameter $H$ (left panel) 
and reduced Hubble parameter $\tilde H=H/(\om h^2)^{1/2}$, i.e.\ Hubble 
parameter relative to high redshift (right panel), are plotted vs.\ 
redshift, leaving spatial curvature free.   The behavior is similar to 
that in Figure~\ref{fig:sensd} but the degeneracies are somewhat more 
severe. 
}
\label{fig:sensh}
\end{center}
\end{figure}

Because distances are integrals over the expansion history, 
which in turn involves an integral over the equation of state, degeneracies 
exist between cosmological parameters contributing to the distance.  
These are common to all distance measurements.  
Figures~\ref{fig:sensd}-\ref{fig:sensh} illustrate the sensitivity and 
degeneracy of 
various expansion history measurements to the cosmological parameters. 
Here $d$ is the luminosity or angular distance (we consider fractional 
precisions so the distinguishing factor of $(1+z)^2$ cancels out), $H$ 
is the Hubble parameter, or expansion rate, and the quantities 
$\tilde d=d\,(\om h^2)^{1/2}$, $\tilde H=H/(\om h^2)^{1/2}$, where $h$ 
is the dimensionless Hubble constant, are relative to high redshift.

First we note that for a given fractional measurement 
precision, distances contain about as much information as the Hubble 
parameter, despite the extra integral (the greater lever arm acts to 
compensate for the integral).  That is (assuming flatness 
for simplicity), 
\beqa 
d&=&\int dz/H, \\ 
\frac{1}{d}\frac{\partial d}{\partial\theta}&=&-\int\frac{dz}{H^2} 
\frac{\partial H}{\partial\theta}\Bigg/\int\frac{dz}{H}= 
-\left\langle\frac{\partial\ln H}{\partial\theta}\right\rangle, 
\eeqa 
for a parameter $\theta$, where angle brackets denote the weighted average.  
However, surveys 
to measure distances to a given precision can be much less 
time consuming than those to measure the Hubble parameter.  

Concerning the physical interpretation of the partial derivatives, 
sometimes known as Fisher sensitivities since they also enter the 
Fisher information matrix \citep{TegFis}, 
recall that 
\beq 
\delta\theta=\delta(\ln d)\Bigg/ 
\left(\frac{\partial\ln d}{\partial\theta}\right). 
\eeq 
If for some parameter the Fisher partial derivative at some redshift is 
0.88, say, then 
a 1\% measurement gives an unmarginalized uncertainty on the parameter 
of 0.011.  If the parameter is $w_a/10.8$, say, this means the 
unmarginalized uncertainty on $w_a$ is 0.12.  The higher the denominator 
in the parameter label, the less sensitivity exists to that parameter.

The greatest sensitivity is to the energy densities in matter and dark 
energy; note that we do not here assume a flat universe.  The nature 
of the dark energy is here described through the equation of state 
parametrization $w(a)=w_0+w_a(1-a)$, where $w_0$ is the present value 
and $w_a$ provides a measure of its time variation (also see 
\S\ref{sec:param}).  This has been shown 
to be an excellent approximation to a wide variety of origins for the 
acceleration \citep{Linprl}.  Sensitivity to $w_a$ is quite modest so 
highly accurate measurements are required to discover the physics 
behind acceleration. 

Note that degeneracies between parameters, e.g.\ $\om$ and $w_a$, and 
$\Omega_w$ and $w_0$, are strong at redshifts $z\la1$.  Not until 
$z>1.5$ do distance measurements make an appreciable distinction between 
these variables.  This determines the need for mapping the expansion 
history over approximately the last e-fold of expansion, i.e.\ to $a=1/e$ 
or $z=1.7$.  Since the flux of distant sources gets redshifted, this 
requires many observations to move into the near infrared, which can only 
be achieved to great accuracy from space. 

From the right panel of Figure~\ref{fig:sensd} we see that the degeneracy 
between the matter density $\om$ and dark energy equation of state time 
variation $w_a$ holds regardless of whether space is assumed flat or not. 
The dark energy density $\Omega_w$ and the dark energy present equation 
of state $w_0$ stay highly degenerate in the curvature free case, whether 
considering $d$ or $\tilde d$ (not shown in the figure). 

From the relation between the derivatives of the distance with respect 
to cosmological parameters, 
as a function of redshift, one can intuit the orientation of confidence 
contours in various planes.  For example, the sensitivity for the matter 
density and dark energy density enter with opposite signs, so increasing 
one can be compensated by increasing the other, explaining the low 
densities to high densities diagonal orientation in the 
$\om$-$\Omega_\Lambda$ plane.  Since the sensitivity to $\om$ trends 
monotonically with redshift, while that of $\Omega_w$ has a different shape, 
this implies that observations over a range of redshifts will rotate 
the confidence contours, breaking the degeneracies and tightening the 
constraints beyond the mere power of added statistics.  Similarly one can 
see that contours in the $\om$-$w_0$ plane will have negative slope, as 
will those in the $w_0$-$w_a$ plane. 

For the tilde variable $\tilde d$, that is distances relative to high 
redshift 
rather than low redshift, there is relatively little degeneracy between 
the matter density and the other variables -- but there is also much 
less sensitivity to the other variables.  The equivalent normalized 
variables to those in the top left panel of Figure~\ref{fig:sensd} are 
$\Omega_w/6.60$, $-w_0/6.59$, and $-w_a/36.7$.  This insensitivity is 
why contours in the $\om$-$\Omega_w$ or $\om$-$w$ planes are rather 
vertical for distances tied to high redshift, like baryon acoustic 
oscillations.  Of course since the degeneracy directions between 
distances tied to low redshift and those tied to high redshift are 
different, these measurements are complementary.  (Although not as 
much in the $w_0$-$w_a$ plane, since high redshift is essentially blind 
to these 
parameters so tilde and regular distances have the same dependence.) 

For the Hubble parameter, a nearly triple degeneracy exists between 
$\om$, $\Omega_w$ and $w_a$ out to $z\approx1$.  Interestingly, for 
$\tilde H$ there is a strong degeneracy between $\om$ and $w_0$ for 
$z\approx1-2$, while the $\Omega_w$-$w_a$ degeneracy remains. 
It is quite difficult to observe $H(z)$ directly, i.e.\ measured 
relative to low redshift; more common is measurement relative 
to high redshift as through baryon acoustic oscillation in the line of 
sight direction (see \S\ref{sec:bao}).  Here, the sensitivity to the 
dark energy equation of state is reduced relative to the 
distance case, even apart from the degeneracies (since $\tilde H$ does 
not include the $z\la1$ lever arm).  

For either $\tilde d$ 
or $\tilde H$ the correlation between energy densities is reversed from 
the regular quantities, so the likelihood contour orientation in the 
$\om$-$\Omega_\Lambda$ plane is reversed as well, close to that of the 
CMB, another measurement tied to the high redshift universe.  The contour 
in $\Omega_m$-$w_0$ is similarly reversed but not in $w_0$-$w_a$.  Thus, 
all these varieties of distance measurements have similar equation of 
state dependencies and, unfortunately, no substantial orthogonality can 
be achieved in this plane. 

\begin{figure}[!htb]
\begin{center}
\psfig{file=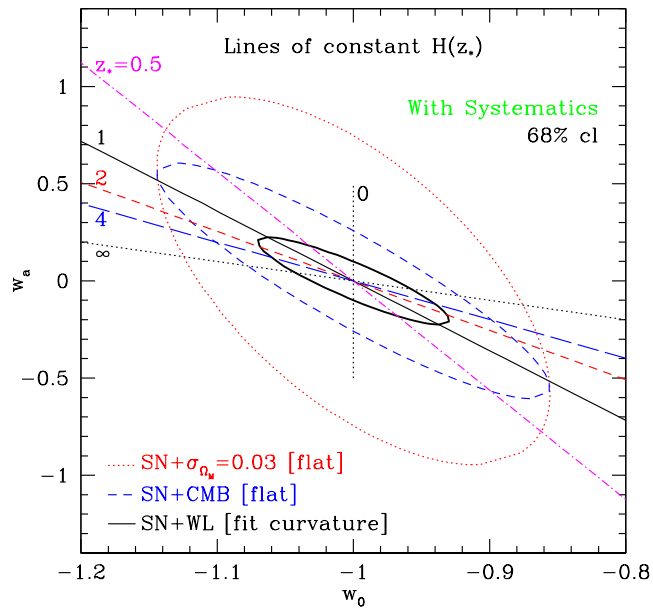,width=3.4in}
\caption{Lines of constant expansion history, i.e.\ Hubble parameter, 
are plotted in the dark energy equation of state plane.  Observations 
over a range of redshifts give complementarity, but never complete 
orthogonality.  Contours show approximate examples of how 
certain combinations of distance probes effectively map different 
parts of the expansion history. 
}
\label{fig:isoww}
\end{center}
\end{figure}

To emphasize this last point, Figure~\ref{fig:isoww} illustrates the 
degeneracy directions for contours of constant Hubble parameter and 
constant matter density at various redshifts.   As we consider redshifts 
running from $z=0$ to $z\gg1$, the isocontours only rotate moderately. 
Since they always remain oriented in the same quadrant, orthogonality 
is not possible.

Conversely, while not giving as much complementarity in the dark energy 
equation of state as we might like, the use of different probes can be 
viewed as giving complementarity in mapping the expansion history. 
Observations of distances out to $z=1.7$, e.g.\ through supernova data, 
plus weak lensing are basically equivalent to mapping the expansion at 
$z\approx1.5$, supernovae plus CMB give it at 
$z\approx0.8$, while supernovae plus a present matter density prior 
provide it at $z\approx0.3$.  

In this subsection we have seen that a considerable part of the optimal 
approach for mapping the expansion 
history is set purely by the innate cosmological dependence, and the 
survey design must follow these foundations as basic science requirements.  
The purpose of detailed design of successful surveys is rather to work 
within this framework to minimize systematic uncertainties in the 
measurements, discussed in more detail in \S\ref{sec:sys}.  These two 
requirements give some of the main conclusions of this review.

\subsection{Type Ia supernovae} \label{sec:sn}

The class of exploding stars called Type Ia supernovae (SN) become as bright 
as a galaxy and can be seen to great distances.  Moreover, while having 
a modest intrinsic scatter in luminosity, they can be further calibrated 
to serve as standardized candles.  Used in this way as distance measures, 
studies of SN led to the discovery of the acceleration of the universe 
\citep{riess98,perl99}. 

As of the beginning of 2008, over 300 SN with measurement quality suitable 
for cosmological constraints had been published, representing the efforts of 
several survey groups.  Finding a SN is merely the first step: the time 
series of flux (the lightcurve), must be measured with high signal to noise 
from before peak flux (maximum light) to a month or more afterward.  This 
must be done for multiple wavelength bands to permit dust and intrinsic 
color corrections.  Additionally, spectroscopy to provide an accurate 
redshift and confirmation that it is a Type Ia supernova must be obtained. 
See \citet{KimPilar} for further observational issues.  (Also see 
\citet{Nugent} for the use of Type II-P supernovae.) 

Once the multiwavelength fits and corrections are carried out, one derives 
the distance, often spoken of in terms of the equivalent magnitude or 
logarithmic flux known as the distance modulus.  The distance-redshift or 
magnitude-redshift relation is referred to as the Hubble diagram.  This, 
or any other derived distance relation, can then be compared to cosmological 
models.  

\begin{figure}[!htb]
\begin{center}
\psfig{file=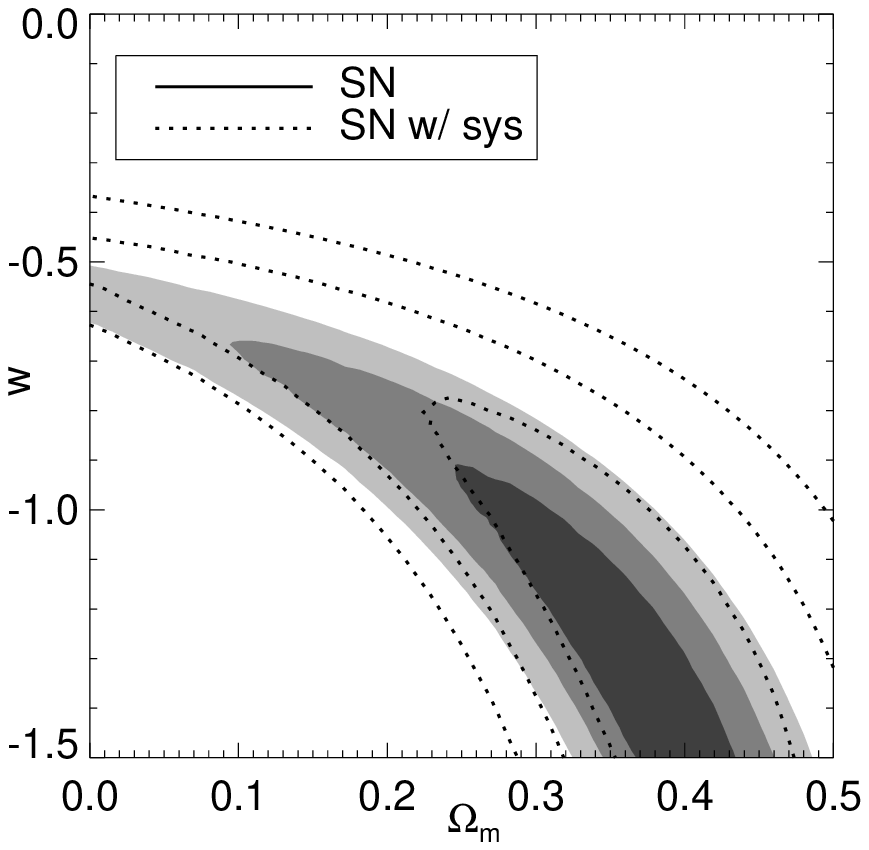,width=3.in}
\psfig{file=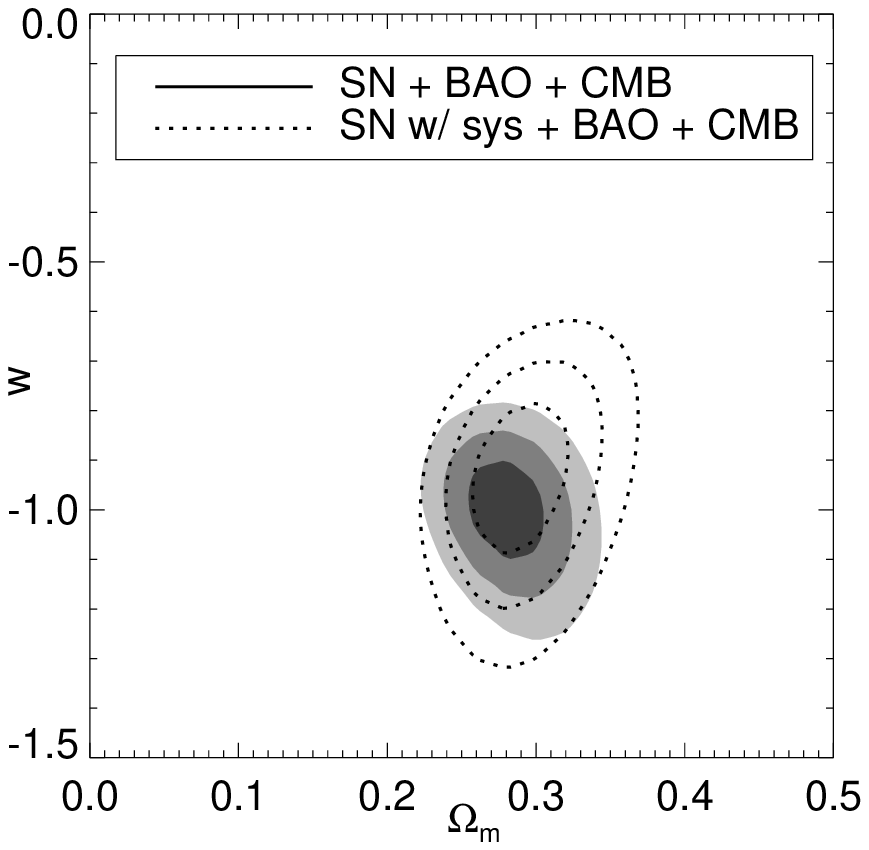,width=3.in}
\caption{68.3\%, 95.4\%, and 99.7\% confidence level contours on the 
matter density and constant equation of state from the Union 2007 set of 
supernova distances.  The left panel shows SN only and the right panel 
includes current CMB and baryon acoustic oscillation constraints. 
From \citet{Kowal}. 
}
\label{fig:kowalw}
\end{center}
\end{figure}

Analysis of the world SN data sets published as of the start of 2008, 
comprising 307 SN, show the expansion history is consistent with a 
cosmological constant plus matter universe, $\Lambda$CDM, but also 
with a great variety of other models \citep{Kowal}.  Essentially no 
constraints on dark energy can be placed at $z>1$ and the limits on 
time variation are no more stringent than the characteristic time 
scale being of the Hubble time or shorter.  Figure~\ref{fig:kowalw} 
shows that even for a flat universe and constant equation of state 
there is considerable latitude in dark energy characteristics and that 
systematic uncertainties need to be reduced for further numbers of SN 
to be useful.

\subsection{Cosmic microwave background} \label{sec:cmb} 

The geometric distance information within the cosmic microwave background 
radiation (CMB) arises from the angular scale of the acoustic peaks in the 
temperature power spectrum, reflecting the sound horizon of perturbations 
in the tightly coupled photon-baryon fluid at the time of recombination. 
If one can predict from atomic physics the sound horizon scale then it 
can be used as a standard ruler, with the angular scale then measuring the 
ratio of the sound horizon $s$ to the distance to CMB last scattering at 
$z_{\rm lss}\approx 1089$.  Related, but not identical to the inverse of 
this distance ratio, $R=\sqrt{\om h^2}\,\dls$ is often called 
the reduced distance or CMB shift parameter.  
This gives a good approximation to the full CMB leverage on the cosmic 
expansion for models near $\Lambda$CDM \citep{EfBond}; for nonstandard 
models where the sound horizon is affected, this needs to be corrected 
\citep{Kowal} or supplemented \citep{Wright}. 

Because the CMB essentially delivers a single distance, it cannot 
effectively break degeneracies between cosmological parameters on its 
own.  While some leverage comes from other aspects than the geometric 
shift factor (such as the integrated Sachs-Wolfe effect), the 
dark energy constraints tend to be weak.  However, as a high redshift 
distance indicator the CMB does provide a long lever arm useful for 
combination with other distance probes (effectively the $z=\infty$ line 
in Figure~\ref{fig:isoww}), and can strongly aid in breaking 
degeneracies \citep{fhlt}. 

\begin{figure}[!htb]
\begin{center}
\psfig{file=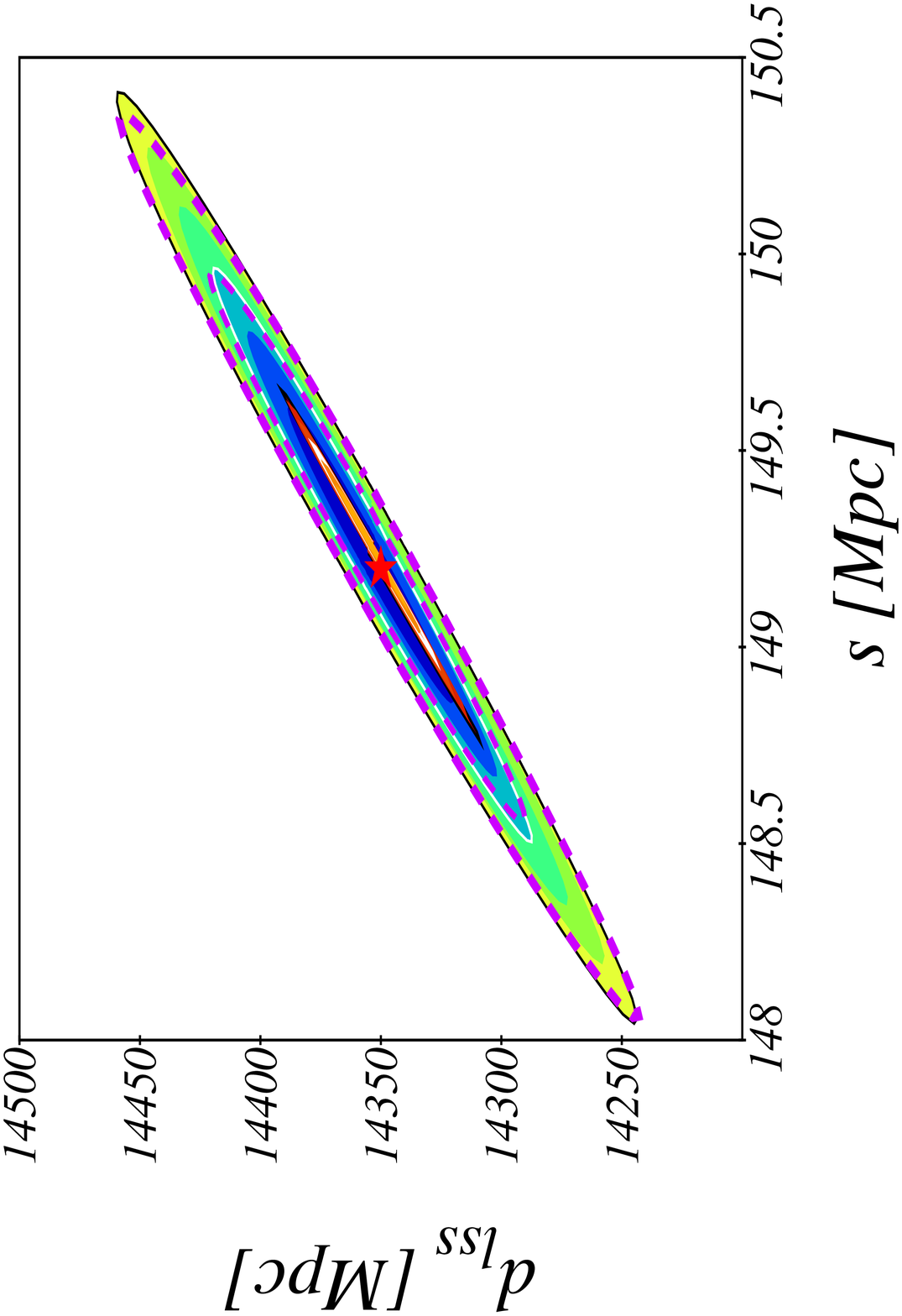,angle=-90,width=3.in}
\psfig{file=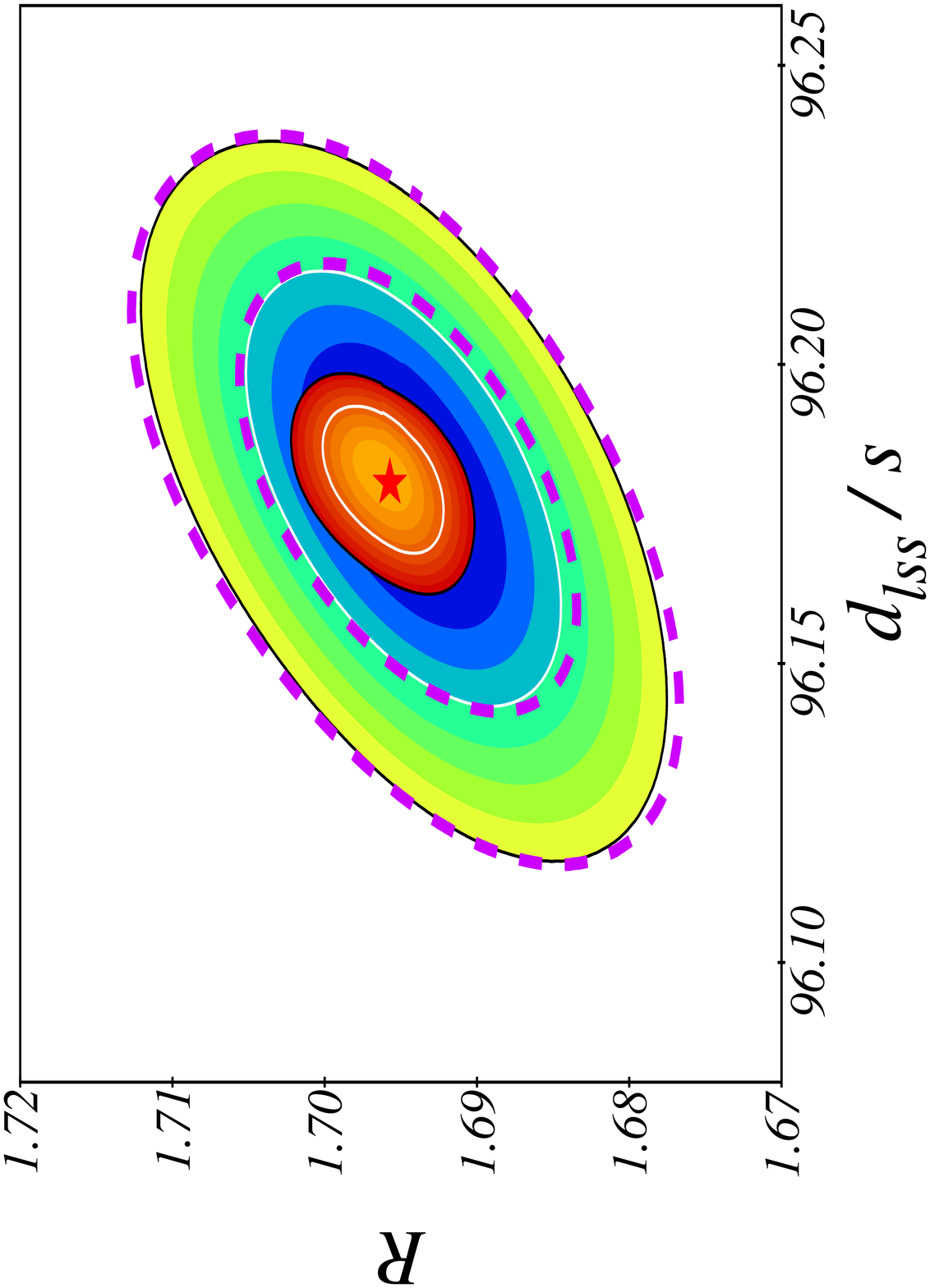,angle=-90,width=3.in}
\caption{{\it Left panel:} CMB data determines the geometric quantities of 
the distance to last scattering $d_{\rm lss}$ and the sound horizon 
scale $s$ precisely, and their ratio (the slope of the contours) -- the 
acoustic peak angle -- even 
more so.  The contours give the 68.3\% (white line) and 95.4\% (black 
line) confidence levels for a cosmic variance limited experiment. 
Outer (dark blue to light green) contours use the temperature power spectrum, 
with dashed contours restricted to multipoles $\ell\ge40$, while 
the inner (light gold to dark red) contours include the E-mode polarization 
and TE cross-spectrum.  {\it Right panel:} Similar contours for the 
shift parameter $R=\sqrt{\om h^2}\,d_{\rm lss}$ and acoustic peak scale. 
}
\label{fig:dlsss}
\end{center}
\end{figure}

The reduced distance to last scattering, being measured with excellent 
precision -- 1.8\% in the case of current data and the possibility of 0.4\% 
precision with Planck data -- defines a thin surface in cosmological 
parameter space (see, e.g., \citet{HuHut3D}).  From Figure~\ref{fig:dlsss} 
we see that the acoustic 
peak structure contains substantial geometric information for mapping the 
cosmological expansion.  This will provide most of the information on 
dark energy since cosmic variance prevents the low multipoles from providing 
substantial leverage (and almost none at all to the geometric quantities, 
as shown by the similarity of dashed and solid contours).  Polarization 
information from $E$-modes and the 
cross spectrum improves the geometric knowledge.  The distance ratio 
$d_{\rm lss}/s$ of the distance to last scattering relative to the sound 
horizon is determined much better (almost 20 times so) than the reduced 
distance to last scattering $R=\sqrt{\om h^2}\,d_{\rm lss}$, but there 
is substantial covariance between these quantities (as evident in the 
right panel) for models near $\Lambda$CDM 
(unless broken through external priors on, e.g., the Hubble constant).

The tightly constrained geometric information means that certain 
combinations of cosmological parameters are 
well determined, but this can actually be a pitfall if one is not careful 
in interpretation.  Under certain circumstances, one such parameter is the 
value of the dark energy equation of state at a particular redshift, 
$w_\star=w(z\approx0.4)$, related to a so-called pivot redshift.  If CMB 
data is consistent with a cosmological constant universe, $\Lambda$CDM, 
then this forces $w_\star\approx-1$.  However, this does not mean that 
dark energy is the cosmological constant -- that is merely a mirage as 
a wide variety of dynamical models would be forced to give the same 
result.  See \S\ref{sec:mirage} for further discussion.

\subsection{Baryon acoustic oscillations} \label{sec:bao} 

The sound horizon imprinted in the density oscillations of the 
photon-baryon fluid, showing up as acoustic peaks and valleys in the 
CMB temperature power spectrum, also appears in the large scale spatial 
distribution of baryons.  Since galaxies, galaxy clusters, and other 
objects containing baryons can be observed at various redshifts, 
measurements of the angular scale defined by the 
standard ruler of the sound horizon provide a distance-redshift probe. 
Table~\ref{tab:baocmb} compares the acoustic oscillations in baryons and 
photons.

\begin{table}[!htb]
\caption{Comparison of signals from perturbations in the prerecombination 
coupled baryon-photon fluid. 
}
\label{tab:baocmb} 
\begin{center}
\begin{tabular}{|l|c|c|} 
\hline
& Photons & Baryons \\ 
\hline 
Effect & CMB acoustic peaks & Baryon acoustic oscillations \\ 
Scale & $1^\circ$ & $100\,h^{-1}$ Mpc comoving \\ 
Base amplitude & $5\times 10^{-5}$ & $10^{-1}$ \\ 
Oscillation amplitude & ${\mathcal O}(1)$ & 5\% \\ 
Detection & $10^{15}$/hand/sec & indirect: light from $<10^{10}$ 
galaxies \\ 
\hline 
\end{tabular}
\end{center}
\end{table}

For spatial density patterns corresponding to baryon acoustic oscillations 
(BAO) transverse to the line of sight, this is an angular distance 
relation while radially, along the line of sight, this is a proper 
distance interval, corresponding in the limit of small interval to the 
inverse Hubble parameter since $dr_{\rm prop}=dz/[(1+z)H]$.  It is 
important to remember that the quantities measured are actually distance 
ratios, i.e.\ $\tilde d=d_a/s$ and $\tilde H=sH$.  These give different 
dependencies on the cosmological parameters than SN distances, as rather 
than being ratios relative to low redshift distances they are ratios 
relative to the high redshift universe (see \S\ref{sec:distgen} and 
Figures~\ref{fig:sensd}-\ref{fig:sensh}). 

The BAO scale was first measured with moderate statistical significance 
in 2005 with Sloan Digital Sky Survey data \citep{eisbao,huetsi,nikhbao} 
and 2dF survey data \citep{twodf}. 
Current precision is 3.6\% on the angular distance measurement at $z=0.35$ 
and 6.5\% at $z=0.5$.  Comparing the $z=0.35$ SDSS result with the 2dF 
result at $z=0.2$ reveals some tension \citep{percivalmay07}, likely 
between the data sets \citep{sanchezcole} rather than from major deviation 
from the $\Lambda$CDM model.  Note that the 
quantities quoted to date are not individual reduced distances but rather 
a roughly spherically averaged quantity convolving angular distance and 
proper distance interval. 

To obtain the proper distance interval, and hence nearly the reduced 
Hubble parameter, requires accurate spectroscopic information to avoid 
projection of modes along the line of sight.  Photometric redshifts 
allow estimation of the reduced angular distance but even that is at 
reduced precision until the redshift resolution approaches spectroscopic 
quality \citep{NikhilBAOz}.  Since the sound horizon scale is about 1/30 
of the Hubble 
radius today, assembling sufficient statistics to measure accurately 
the subtle signal requires huge numbers of baryon markers over great 
volumes ($10^6$-$10^9$ over 1-100 Gpc$^3$).  
For early papers on the cosmological use of BAO and issues 
regarding the interpretation of data, see 
\citet{Meiksin,Cooconf01,eis02conf,glazeblake03,linbao03,huhaiman,eisseo03,blakeglaze05,whitebao05,linbao05,seoeis05}. 

\subsection{Other methods} \label{sec:othdist} 

Many other probes of the expansion history and cosmic acceleration have 
been suggested, in the geometry, geometry+mass, or geometry+mass+gas 
categories.  However many of them either lack a robust physical foundation 
or have yet to achieve practical demonstration of cosmological constraints. 
We therefore merely list some of the geometric methods with a few brief 
comments. 

{\it General tests of the expansion\/} -- These are in a certain sense 
the most fundamental and interesting.  One well 
known test is that the CMB temperature should evolve with redshift as 
$T\sim 1+z$.  This depends on the phase space density of the radiation 
being conserved, following Liouville's Theorem, e.g.\ photons do not 
convert into axions or appreciably interact with other components not in 
equilibrium with the radiation.  It also requires the expansion to be 
adiabatic.  Measurements to date, out to $z=3$, show agreement with 
$T\sim 1+z$ at modest precision, and further prospects include 
using the Sunyaev-Zel'dovich effect (Compton upscattering) 
in clusters to sample the radiation field seen by galaxy clusters at 
different redshifts \citet{ClusCooray}. 

Another test depending on Liouville conservation is the thermodynamic, 
or reciprocity, relation that can be phrased as a redshift scaling 
between angular distances and luminosity distances or ``third party'' 
angular distances between points not including the observer.  Most 
commonly the relation is written as $d_l=(1+z)^2 d_a$.  This has a 
long history in cosmology, dating from the 1930s with Tolman \citet{Tolman}, 
Ruse \citet{Ruse}, 
and Etherington \citet{Etherington}, through the 1970s with Weinberg 
\citet{Wbg72}, 
and a general 
proof in terms of the Raychaudhuri equation and the second law of 
thermodynamics by \citet{Ellis71,Lin88ra}.  The thermodynamic connection is 
basically that if two identical blackbodies sent photons to each other 
in a cyclic process 
over cosmic distances then work would be done unless the relation holds. 

While the generality of these two probes as tests of the nature of the 
expansion itself is intriguing, if a violation were found one might be 
far more likely to blame uncertainties in the sources than a radical 
breakdown of the physics.  

{\it Epochs of the expansion\/} -- The expansion history can also be 
probed at certain special epochs.  For example, light element abundances 
are sensitive to the expansion rate during primordial nucleosynthesis 
\citep{Kaplinghat}.  The CMB recombination epoch and sound horizon 
are influenced by deviations from matter and radiation dominated 
expansion \citep{ZahnZal,ChanChu}.  In the future, if we measure the 
abundance and mass of dark matter particles as thermal relics, we may 
be able to constrain the expansion rate at their freezeout epoch near 
perhaps 1 TeV ($z=10^{16}$) \citep{ChungDSU}.  Such early constraints are 
particularly interesting in multifield or multiepoch acceleration models 
\citep{Griest}.  

{\it Gamma ray bursts\/} -- As sources detectable to high redshift and 
insensitive to dust extinction, some hope existed that gamma ray bursts 
could serve as standardized candles to high redshift (but see 
\S\ref{sec:distsum} and Figure~\ref{fig:mzcut} regarding the limited 
leverage of high redshift).  However, while SN have an intrinsic flux 
dispersion of less than 50\% that one calibrates down to 15\% scatter, 
GRBs start with an isotropic-equivalent energy dispersion of a factor 1000, 
making any calibration much more challenging.  
Standardization relations were 
often ill defined and sensitive to environment \citep{bloomfried} 
and with better data it is now realized that much of the correlation arose 
from experimental selection effects \citep{ButlerApJ671656}. 

{\it Gravitational wave sources\/} -- The intrinsic amplitude of clean systems 
like inspiraling black hole binaries can be predicted from observations 
of the gravitational waveform, giving a standard 
siren to compare to detected gravitational wave amplitude, thus providing 
a distance \citep{schutz86,holzhughes04}.  No such observations have yet 
been achieved.  Obtaining the redshift part of the distance-redshift 
relation depends on observing an electromagnetic counterpart. 

{\it Gravitational lensing cosmography\/} -- Deflection of light by 
massive objects depends on the mass distribution, the distances of 
source and lens from the observer, and the laws of gravity.  However, 
cross-correlation techniques between sources at different redshifts have 
been proposed to reduce the need for knowing the mass distribution 
to extract distances \citep{TaylorJain,BernsteinJain,ZhangSteb}. 
Practical demonstration has not yet occurred, but is an exciting prospect, 
and gravitational lensing also serves as a growth probe (see 
\S\ref{sec:lens44}). 

{\it Environment dependent methods\/} -- Techniques that intrinsically 
depend on not just a single source but its environment or group properties 
require greater understanding of many astrophysical factors.  Such 
techniques often have 
not yet established a clear or robust physical motivation for the 
standardization essential to a distance probe.  Approaches attempted 
include the galaxy age-redshift relation 
\citep{JimenApJ593622,Jimen0412269,Staro07052776}, active galactic nuclei 
radio jets \citep{Daly07105112}, active galactic nuclei 
reverberation mapping \citep{reverb07051722,Teer0510382}, and 
starburst galaxies \citep{Starburst0410612}.

\subsection{Summary of mapping techniques} \label{sec:distsum} 

\begin{itemize} 

\item To map the cosmological expansion accurately requires clear, robust, 
mature measurement techniques.  

\item Observations over a range of redshifts 
are required to break degeneracies, or equivalently rotate likelihood 
contours.  Complementary probes do this as well.  

\item Systematics control can 
narrow contours and prevent biases in parameter estimation; see 
\S\ref{sec:sys} for further analysis and discussion. 

\end{itemize} 

We have seen that much of the leverage for mapping the cosmic expansion 
is determined by the innate cosmology dependence on the parameters, and 
any survey design must work within this framework.  One implication of 
this is that there is a survey depth of diminishing returns; beyond some 
redshift $z\approx2$ little additional leverage accrues to measuring 
such high redshift distances relative to low redshift distances.  That is, 
although the lever arm is longer, the lever is weaker.  For high redshift 
distances measured relative to the high redshift universe, the lever arm 
is shorter, so again improvement in constraints stalls at high redshift. 

\begin{figure}[!htb]
\begin{center}
\psfig{file=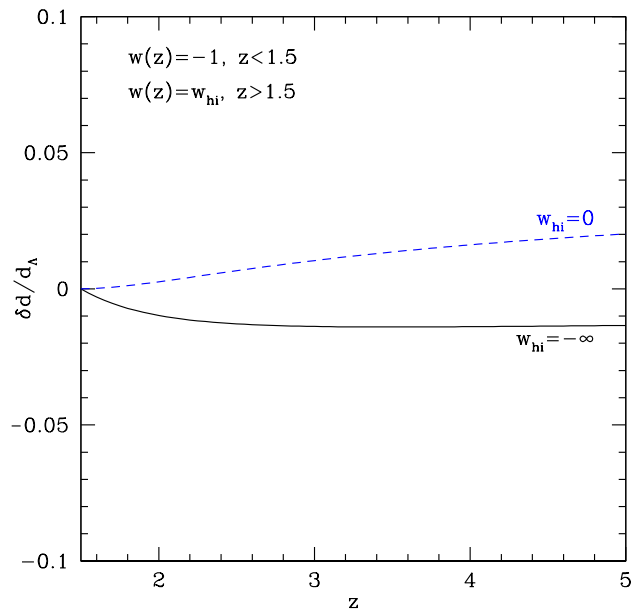,width=3.4in}
\caption{Once the most recent two-thirds of the cosmological expansion 
is determined, deeper measurements of distances tied to low redshift 
have little further leverage.  Over the physically viable range for 
the high redshift dark energy equation of state $w\in[-\infty,0]$, 
distances do not vary between models by more than 1-2\%, exceedingly 
difficult to measure accurately at high redshift.  Cosmological 
considerations alone indicate that covering the range out to $z=1.5-2$ 
is the optimal strategy. 
}
\label{fig:mzcut}
\end{center}
\end{figure}

Figure~\ref{fig:mzcut} illustrates that once the expansion history is 
determined out to, say, $z\approx1.5$ (or 2/3 the age of the universe) that 
it is exceedingly difficult to measure geometrically more about dark energy 
with any 
significance.  Even the simple question of whether the equation of state 
has {\it any\/} deviation from the cosmological constant requires percent 
level accuracy.  Note that at high redshift the range $w\in[-\infty,0]$ is 
comprehensive in that $w=-\infty$ means the dark energy density 
contributes not at all, and $w>0$ would mean it upsets matter domination 
at high redshift (limits on this are discussed further in 
\S\ref{sec:lingro}).  Thus distance probes of the expansion history 
are driven naturally by cosmology to cover the range from $z=0$ to 
1.5--2.

\section{Growth and expansion} \label{sec:gro}

Within the theory of general relativity, the cosmic expansion history 
completely determines the growth of matter density perturbations on 
large scales.  On smaller scales the state of the matter, e.g.\ the Jeans 
length defined by pressure effects, enters as well and when density 
fluctuations go nonlinear then radiative and gas heating and cooling, star 
formation, feedback etc.\ affect the formation and evolution of structure. 
Here we consider only the large scale, linear regime and the role that 
growth observations can play in mapping the cosmological expansion. 

\subsection{Growth of density perturbations} \label{sec:lingro}

For a matter density perturbation $\delta\rho$, its evolution is reduced 
from the Newtonian exponential growth on the gravitational dynamical 
timescale $t_{\rm dyn}=1/\sqrt{G\rho}$ to power law growth on a Hubble 
timescale by the drag induced from the cosmological expansion.  Generally, 
\beq 
g''+\left[4+\frac{1}{2}(\ln H^2)'\right]g' 
+\left[3+\frac{1}{2}(\ln H^2)'-\frac{3}{2}\om(a) \right]g=0, \label{eq:groh} 
\eeq 
where $g=(\delta\rho/\rho)/a$ is the normalized growth, a prime 
denotes differentiation with respect to $\ln a$, and the dimensionless 
matter density $\om(a)=\om a^{-3}/[H/H_0]^2$.  Thus the growth indeed 
depends only on the expansion history $H$ (the matter density is 
implicit within the high redshift, matter dominated expansion behavior). 

In a matter dominated epoch, the solution is $\delta\rho/\rho\sim a$, 
so we defined $g$ such that it would be constant (which we can arrange to 
be unity) in such a case.  In an epoch dominated by an unclustered 
component with equation of state $w$, matter density perturbations do 
not grow.  Note that this is due to the small source term, not the Hubble 
drag -- indeed the Hubble drag term is proportional to $5-3w$ and is less 
strong in a radiation dominated universe than a matter dominated universe, 
yet matter density perturbations still do not grow in the radiation epoch. 
If one somehow fixed $\om(a)$ as one changed $w$, then increasing $w$ 
indeed aids growth.   For domination by a component with $w<0$ (not 
acceleration per se) 
there is the double whammy of reduced source term and increased friction. 

As the universe makes the transition from being dominated by a $w>0$ 
component to matter domination, the solution is 
\beq 
\delta\rho/\rho\sim 1+\frac{\om}{\Omega_w}\frac{1}{w(1+3w)}\,a^{3w}. 
\eeq 
One can see the transition from no growth $\delta\sim{\rm constant}$ to 
growth (with $\delta\sim a$ for a radiation-matter transition). 
For a transition from matter domination to being dominated by a $w<0$ 
component, to first order 
\beq 
\delta\rho/\rho\sim a 
\left[1-\frac{1-w}{-w(5-6w)}\frac{\Omega_w}{\om}\,a^{-3w}\right]. 
\eeq 

Finally, for a $w=0$ component that does not behave in the standard 
fashion, e.g.\ a constant fraction $F$ of the matter density does not 
clump, then 
\beq 
\delta\rho/\rho\sim a^{(\sqrt{25-24F}-1) 
/4}. \label{eq:gro24} 
\eeq 
For example, a small fraction $\Omega_e$ of early dark energy with $w=0$ 
leads to $\delta\rho/\rho\sim a^{1-(3/5)\Omega_e}$ at early times.  
See \citet{fry85,linpert88,fpoc} for early discussion of these 
growth behaviors in multicomponent universes. 

Since 
the origin of the density perturbation source term is the Poisson equation, 
basically contributing $4\pi G_N\delta\rho$, any modification to this 
equation gives an effective $F(k,a)\equiv[G\om(k,a)]/[G_N\om(a)]_{\rm std}$ 
that produces modified growth, where the wavenumber $k$ allows for spatially 
dependent modifications.  If $F$ is nearly constant in time, then the growth 
is determined by Eq.~(\ref{eq:gro24}).  However, if 
the physical origin of the modification also affects the expansion, as from 
a time dependent gravitational coupling $G_N$, then one must alter the 
other growth equation terms as well to obtain the solution. 

\begin{figure}[!htb]
\begin{center}
\psfig{file=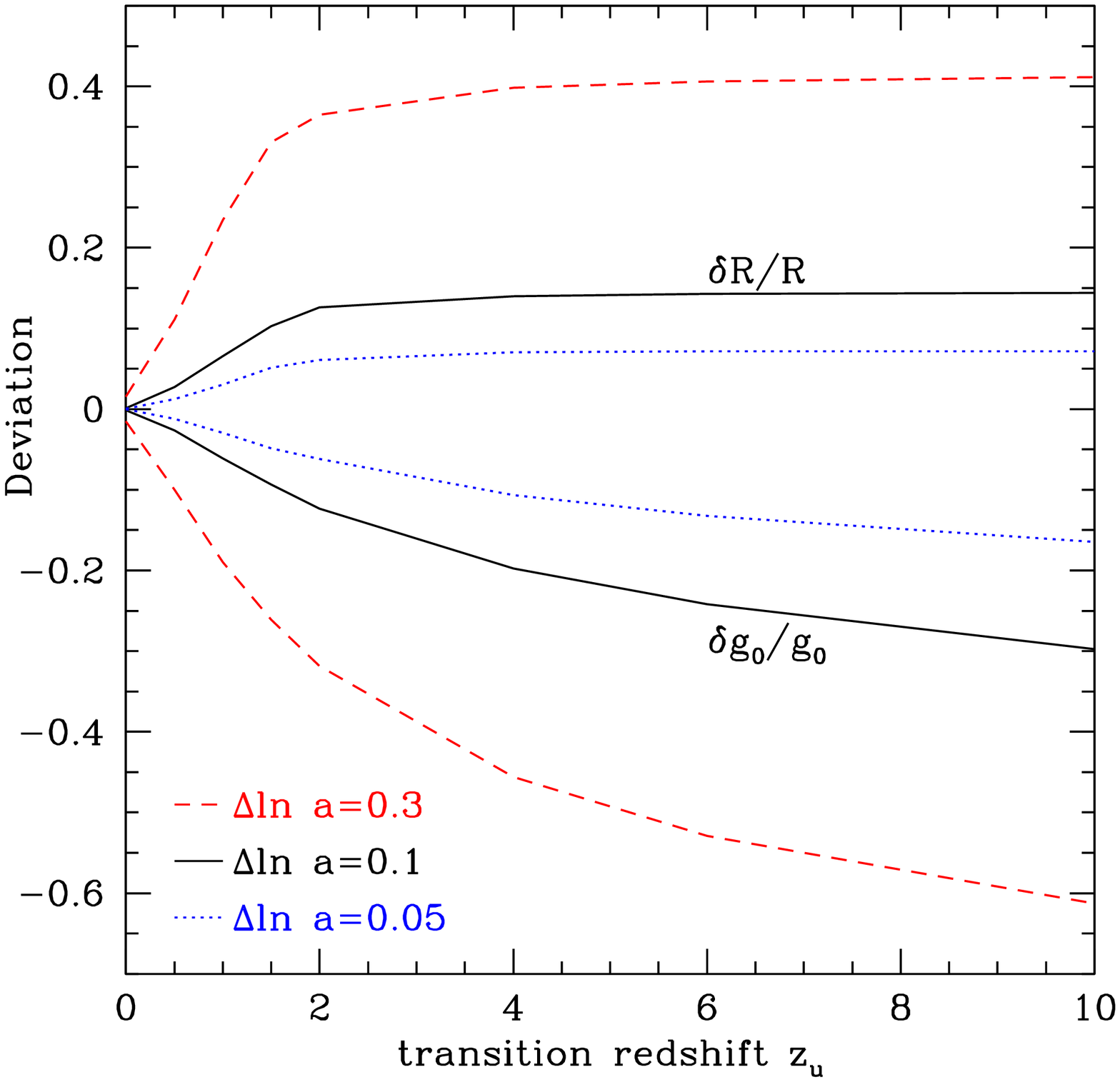,width=3.in}
\psfig{file=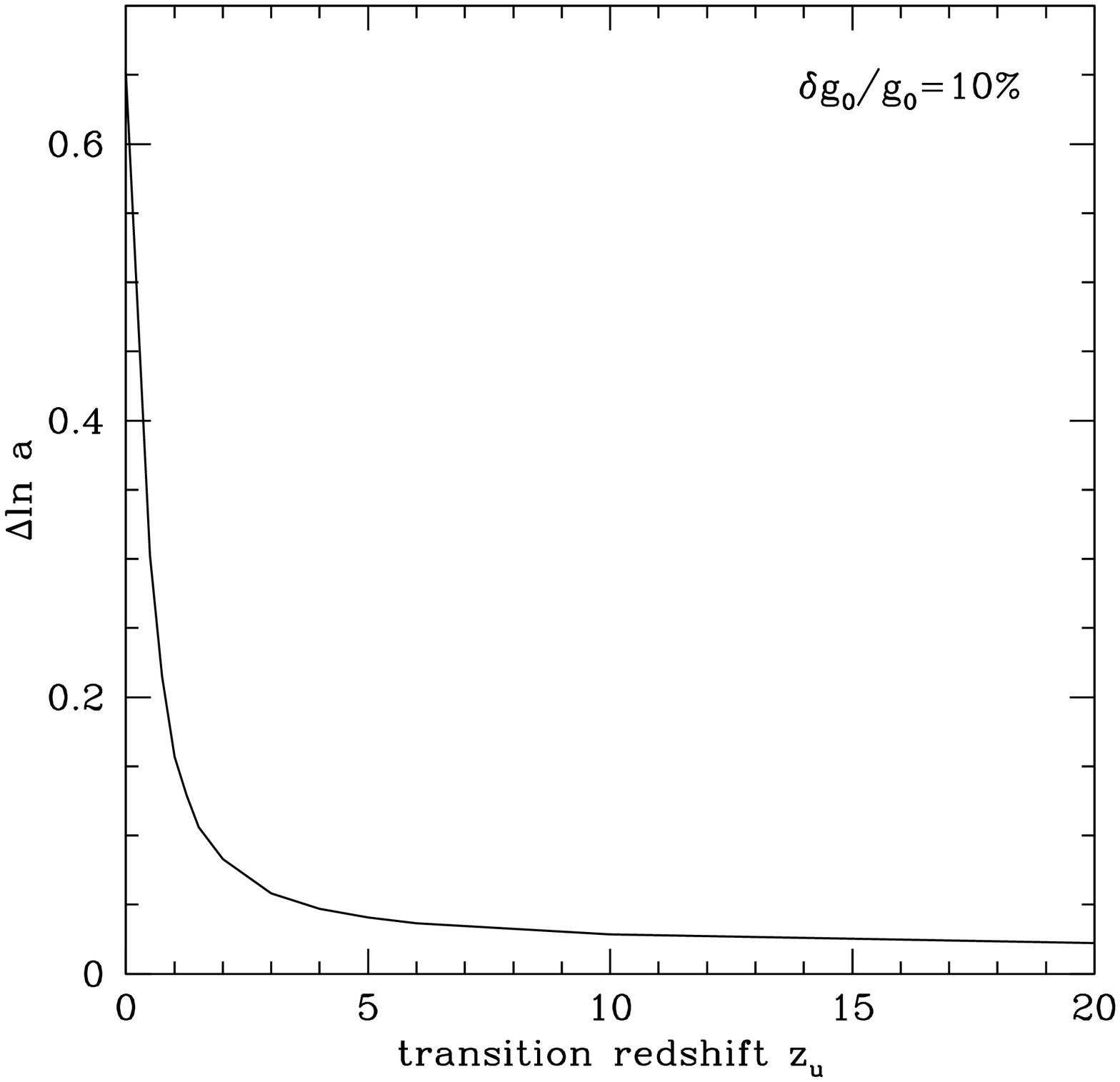,width=3.in}
\caption{Growth measurements can strongly constrain events in the 
cosmological expansion such as an intermediate epoch of acceleration. 
The left panel show the deviations from the $\Lambda$CDM growth, in the 
present total growth $g_0$ and growth ratio $R=g(a=0.35)/g_0$, for a 
transition to total equation of state $-1$ lasting a period 
$\Delta\ln a$, ending at redshift $z_u$.  The right panel shows the 
limit on the length of the acceleration allowed as a function of 
transition redshift, in order for deviations to the total growth to be 
less than 10\%. 
}
\label{fig:boxacc}
\end{center}
\end{figure}

Growth measurements, being integrals from high redshift, can probe the 
high redshift universe.  For example, a measurement of the growth factor 
at, say, $z=2$ that agrees with a $\Lambda$CDM model mapped out at lower 
redshift ensures that the high redshift epoch of matter domination 
occurred substantially as expected.  Specifically, if the linear growth 
factor at $z=2$ deviates by less than 5\%, then for a monotonically 
varying dark energy model 
described by $w(a)=w_0+w_a(1-a)$, we can limit $w_a<0.6$ or $w(z=2)<-0.6$. 
We can also constrain an intermediate, transient epoch 
of acceleration in the cosmic expansion.  Growth measurements can place 
tight constraints on this mechanism for easing the coincidence problem 
\citep{LinEarly}, as shown in Figure~\ref{fig:boxacc}.  To prevent strong 
deviations in growth that would be evident in measurements, the period of 
such acceleration must be much shorter than the characteristic Hubble 
timescale (and hence apparently ``unnatural'').

Overall, the growth history has the potential to constrain the 
expansion history.  However, there are some obstacles.   The growth also 
depends on the initial conditions, couplings to other components, and 
deviations of gravity from general relativity.  For example, in a 
radiation dominated universe growth should not occur, but if the 
perturbation evolution has some ``velocity'' from a previous matter 
dominated epoch it can proceed with a slow, $\ln a$ growth.  (Similarly, 
growth persists into an accelerating epoch.)  Another 
example is that in the prerecombination universe baryons were tightly 
coupled to photons, preventing growth \citep{Meszaros}.  Finally, it is 
quite difficult to detect the matter perturbations per se, instead we 
observe light from galaxies -- a biased tracer of the density field.  
Separating the absolute growth factor from the bias, or the relative 
growth from an evolution in bias, is nontrivial \citep{PhysRepBias}. 
Weak gravitational lensing (see \S\ref{sec:lens44} below) offers a 
way around these issues. 

Measurements of growth are presently not clean probes of the 
expansion history, especially since, as Eq.~(\ref{eq:groh}) indicates, 
precision knowledge of $\om$ is required to truly use this approach to 
map $H(z)$.  Redshift distortion measurements of the growth of the 
matter velocity field (rather than density), employing $d\ln g/d\ln a$, 
have similar issues.  See \citet{Teg06,MacdonLya} for good overviews of 
the current status of growth measurements.

\subsection{Abundance tests} \label{sec:abund} 

Under gravitational instability, mass aggregates and galaxies and clusters 
of galaxies form, evolving in number and mass.  The abundance as a 
function of redshift of particular classes of objects will involve 
(among other variables) the initial power spectrum, astrophysical 
processes such as dissipation and feedback, and the expansion rate of 
the universe, as well as a robust observational proxy for mass.  Two 
famous examples of using galaxy 
abundances as cosmological probes are in a 1975 paper ``An Accelerating 
Universe'' \citep{GunnTins} and in 1986 ``Measurement of the Mass Density 
of the Universe'' \citep{LohSpil}. 
As the titles suggest, the data led to opposite conclusions and each was 
quickly recognized to be showing more about astrophysical evolution of 
the objects than the cosmological expansion history. 

New generations of experiments are underway to count sources as a function 
of redshift and mass, selecting samples through their detection in the 
Sunyaev-Zel'dovich effects, X-ray flux, optical flux, or weak gravitational 
lensing shear.  Currently, data is insufficient to give robust cosmological 
constraints on the expansion history. 
Note that observations do not actually measure an abundance per comoving 
volume, $dN(m)/dV$, which would show how structures grow in mass while 
their (nonevolving) numbers should stay constant.  Rather they see $dN/dz$ 
which involves distances through an extra factor of $dV/dz$.  Thus 
observed abundances do not give a pure growth probe but mix geometry and 
growth -- both an advantage and disadvantage.

\subsection{Gravitational lensing} \label{sec:lens44} 

The gravitational potentials of massive structures deflect, focus, and shear 
bundles of light rays from more distant sources.  The particular shear 
pattern combines information from both the lensing mass and the ``focal 
length'' defined by the source and lens distances from the observer. 
To date, most such data provides leverage on a combination of the matter 
density $\om$ and the rms amplitude of mass fluctuations $\sigma_8$. 
A few surveys so far quote further cosmological constraints -- on the dark 
energy parameters in combination with other data sets \citep{JarvisB}, or 
on the presence of growth \citep{Benjamin100,CFHLSdeep,Massey,Fu,DoreHoek}. 

A number of interesting concepts for using gravitational lensing to map 
the cosmological expansion exist, though practicality has yet to be 
demonstrated.  Both the magnification or convergence fields and the shear 
field carry cosmological information.  For use as a geometric probe, one 
must separate astrophysical or instrumental effects, deconvolve the 
lens gravitational potential model, and measure the redshifts or redshift 
distribution of the lenses and sources accurately.  See 
\citet{TaylorJain,BernsteinJain,ZhangSteb} for discussion of weak lensing 
as a geometric probe.  Similarly, lensing offers promise to probe the 
growth history.  Lensing of the CMB, where the source redshift is known, 
is another interesting application \citep{CMBlens}.  See \citet{ARAAWL} 
for far more discussion of weak lensing and its many uses than given here.

\subsection{Testing gravity} 

\subsubsection{First steps} 

Although the focus of this article is on mapping the cosmological 
expansion, since the growth history depends on both the expansion and 
the laws of gravity one could consider using expansion plus growth 
measurements to test the gravitational framework.  This is an exciting 
possibility that has recently received considerable attention in the 
literature (see, e.g., \citet{LueSS,Lin04,KnoxST,IshakUS} for early 
work), although current data cannot provide significant constraints.  

In addition to comparing predictions of specific theories of gravity to 
data, one basic approach is parametrization of gravitational cosmological 
effects.  While one could define cosmological parameters $\om^{\rm gr}$, 
$w_0^{\rm gr}$, $w_a^{\rm gr}$, say, derived from fitting the growth 
history and 
contrast these with those derived from pure expansion measurements, 
one could also keep the physics of cosmic expansion expressed through 
the well-defined (effective) parameters $\om$, $w_0$, $w_a$ and look 
for physical effects from gravity beyond Einstein relativity in separate, 
clearly interpretable gravitational parameters. 

The gravitational growth index $\gamma$ was designed specifically to 
preserve this distinction of physical phenomena \citep{GroExp}.  Here the 
matter density linear growth factor $g=(\delta\rho/\rho)/a$ is written 
\beq 
g(a)=e^{\int_0^a (da'/a')\,[\om(a')^\gamma-1]}. 
\eeq 
See \citet{GroExp,HutLin06,LinCahn} for discussion of the accuracy, 
robustness, and basis of this form.  However, gravity does more than 
affect the growth: it affects the light deflection law in lensing 
and the relation between the matter density and velocity fields.  
Since two potentials enter generically, in the time-time and space-space 
terms of the metric, 
\citet{CaldCooMel,ZhangLBD,AmenKS,HuSaw,AminWag,ZhangJain,Edbert,Caldslip} 
and others have proposed two or more parameters for testing extensions 
to general relativity 
(GR).  These may involve the ratio of metric potentials (unity for GR), as 
a cosmological generalization of the parametrized post-Newtonian 
quantity, and their difference (zero for GR), related to the anisotropic 
stress.

\subsubsection{Problems parametrizing beyond-Einstein gravity \ } 

The two gravitational parameter approach is an exciting idea that deserves 
active exploration.  Here, however, to balance the literature we 
undertake a more critical overview of this program.   Difficulties 
to overcome include: 

\begin{enumerate} 

\item Reduction of two functions to two (or a small number) of parameters; 

\item Spatial dependence, e.g.\ strong coupling or Vainshtein scales, or 
quantities living in real space vs.\ Fourier space; 

\item Covariance among deviations in gravitational coupling, metric 
potentials, and fluid anisotropic stress. 

\end{enumerate}

We give very brief assessments of each item.  For the first point, a 
successful proof of concept is the gravitational growth index, which 
for a wide class of theories adds a single (constant) parameter to 
describing linear growth.  \citet{LinCahn} explains why this works under 
certain conditions, basically requiring small deviations from 
standard cosmological expansion and gravity.  Many theories indeed 
have small deviations, but these are often so small they become 
problematic to detect; other 
theories with larger deviations are either ruled out by, e.g., solar 
system tests or have scale dependence requiring multiple parameters. 

Regarding the second point, in the parametrized post-Newtonian (PPN) 
formalism the ratio of metric 
potentials is defined in real space, e.g.\ $\Phi(r)/\Psi(r)$.  This 
is not equivalent to defining a parameter $\eta=\Phi_k/\Psi_k$ in 
Fourier space.  Given some function $f(r)=\Phi(r)+\Psi(r)$, it is 
not generically true that a nonlinear function $F(f(r))$ depends in 
the same manner on $\Phi$ and $\Psi$.  Thus many power spectra, e.g.\ 
involving $\langle(\Phi+\Psi)^2\rangle$ for the integrated Sachs-Wolfe 
effect, will not follow the simple formalism. 

Spatial variation also does not preserve the functional dependence; 
for example, while the light deflection angle depends on the specific 
combination $\Phi(r)+\Psi(r)$, the actual deflection at impact parameter 
$b$ is not a function only of $\Phi(b)+\Psi(b)$.  Consider lensing in 
DGP \citep{dgp} gravity, where 
\beqa 
\Phi(r)&=&\frac{m}{r}+\frac{n_\phi}{r^3} \\ 
\Psi(r)&=&\frac{m}{r}+\frac{n_\psi}{r^3} 
\eeqa 
where the $n$'s are related to extradimensional quantities.  
(Many IR modifications of gravity have this form, but note it is very 
different from the Yukawa form so general parametrization of spatial 
dependence may be difficult).  
Thus $f(r)$ takes the form $2m/r+n/r^3$ and some function, say, 
\beq 
F(f(r_0))=\int_{r_0}^\infty \frac{dr}{r}f(r)=\frac{2m}{r_0}+\frac{n}{3r_0^3} 
\ne C\,[\Phi(r_0)+\Psi(r_0)], 
\eeq 
for any constant $C$.  More concretely, for deflection of light at impact 
parameter $b$ from a point mass (extending \citet{KeetonPetters06}), we 
find the deflection angle 
\beq 
\alpha_b=2\,[\Phi+\Psi]_b-6\,[\Psi-\Phi]_b, 
\eeq 
breaking the functional form $\Phi+\Psi$ with an extra term looking like 
a spurious anisotropic stress. 
Thus observations of lensing deflection do not tell us about a simple 
parameter $\eta$ in the form $\Phi\,(1+\eta)$, as we might hope. 

Another issue concerns interdependence among different aspects of 
gravitational modifications.  Variation in gravitational coupling may 
well occur alongside anisotropic stress, as in scalar-tensor theories. 
In Poisson's equation, therefore, one cannot define a Fourier space mass 
power spectrum $\langle\delta_k^2\rangle$ because the physical source 
involves $\langle(G_{\rm eff}\delta_k)^2\rangle$ and it may not be clear 
how the interaction of these two quantities affects the result.  This 
is reminiscent of varying constant 
theories where one can calculate the effects of a varying fine structure 
constant, say, but does not have a unified framework for understanding 
how the electron-proton mass ratio, say, varies simultaneously, and hence 
affects observations as well \citep{UzanRMP}.  Confusion also 
arises between gravitational modifications and fluid microphysics, e.g.\ 
anisotropic stress of a component, coupling between components, or finite 
sound speed (see, e.g., \citet{KunzStress}). 

\subsubsection{Levels of discovery} \label{sec:gravsys}

In summary, there is a rich array of challenges for a program seeking to 
parameterize beyond Einstein gravity.  One ray of hope is that 
different observations depend differently on the quantities, some of 
which are outlined in Table~\ref{tab:thysys} (also see \citet{ZhangJain}).  
In the theory realm, SN serve as pure 
geometric probes of the cosmological expansion, 
immune to the uncertainties imposed by these additional parameters 
(even $G(t)$, see e.g.\ \citet{sndotg}), 
providing the most unambiguous understanding.  This point is worth 
considering: all other methods in use rely on understanding 
the rest of the dark sector -- dark matter and gravity -- as they seek 
to explore dark energy.  As well though, 
we should keep in mind the insight by Richard Feynman: 

\begin{quote} 
``Yesterday's sensation is today's calibration, and tomorrow's background.'' 
\end{quote} 

Mapping the cosmological expansion with a simple, robust, and geometric 
method like supernovae provides a firm foundation as we also must 
probe deeper, with 
techniques that have more complex -- richer -- dependence 
on further variables revealing the microphysics and testing gravity. 
Mutual support among methods will be key to yielding understanding.

\begin{table}[!t]
  \caption{Theory systematics occur even for some geometric probes, the 
cleanest astrophysically.  For example, light deflection depends on 
the metric potentials as $\Phi+\Psi$, plus separating out the mass model, 
and baryon acoustic oscillations are sensitive to the validity of the 
usual CDM growth, 
hence affected by $\Phi$ and $\Psi$ (see, for example, \citet{sawicki} for 
difficulties in the braneworld case), new density perturbations due 
to the sound speed $c_s$ (see, e.g., \citet{dedeo}) and anisotropic 
stress $\pi_s$, variation of the gravitational coupling $G$ with scale and 
time, and altered fluctuations due to matter coupling $\Gamma$ (see, e.g., 
\citet{amendola,mangano}).  
}
  \label{tab:thysys}
  \begin{indented} \item[]
  \begin{tabular}[t]{|c|c|c|}
    \hline%\hline 
    Probe & Theory Systematic (dominant) & Theory Systematic (potential) \\ 
    \hline\hline  
    SN Ia & --- & --- \\ \hline
    WL & $\Phi+\Psi$ & $c_s$, $\pi_s$, $G(k,t)$ \\ \hline 
    BAO & $\Phi$, $\Psi$, $c_s$, $\Gamma$ & 
$\pi_s$, $G(k,t)$ \\ \hline 
  \end{tabular}
  \end{indented} 
\end{table}

\section{Systematics in data and theory} \label{sec:sys}

Accurate mapping of the cosmological expansion requires challenging 
observations over a broad redshift range with precision measurements. 
Technological developments such as large format CCDs, large telescopes, 
and new wavelength windows make such surveys possible.  In addition to 
obtaining large data sets of measurements, we must also address systematic 
uncertainties in measurements and astrophysical source properties.  
Beyond that, for accurate mapping we must consider systematics in the 
theoretical interpretation and the data analysis.  Here we present some 
brief illustrations of the impact of such systematics on mapping the 
expansion. 

\subsection{Parameterizing dark energy} \label{sec:param} 

In extracting cosmological parameters from the data, one wants parameters 
that are physically revealing, that can be fit precisely, and that 
accurately portray the key physical properties.  For exploration of many 
different parametrizations and approaches see \citet{LinGRG} and references 
therein.  Any functional form for the expansion, e.g.\ the dark energy 
equation of state $w(z)$, runs the risk of limiting or biasing the physical 
interpretation, so one can also consider approaches such as binned 
equations of state, defined as tophats in redshift, say, or decomposition 
into orthogonal basis functions or principal component analysis (see, e.g., 
\citet{HutStark}).  However, for a finite number of components or modes 
this does not solve all the problems, and introduces some new ones such as 
model dependence and uncertain signal-to-noise criteria (see 
\citet{dePLinPCA} for detailed discussion of these methods). 

Indeed, even next generation data restricts the number of well-fit 
parameters for the dark energy equation of state to two \citep{LinHut05}, 
greatly reducing the flexibility of basis function expansion or bins.  
Here, we concentrate on a few comments regarding two parameter functions 
and how to extract clear, robust physics from them. 

For understanding physics, two key properties related to the expansion 
history and the nature of acceleration are the value of the dark energy 
equation of state and its time variation $w'=dw/d\ln a$.  These can be 
viewed as analogous to the power spectral tilt and running of inflation.  
The two parameter form 
\beq 
w(a)=w_0+w_a(1-a), 
\eeq 
based on a physical foundation by \citet{Linprl}, has been shown to be 
robust, precise, bounded, and widely applicable, able to match accurately 
a great variety of dark energy physics.  See \citet{wcon} for tests of 
its limits of physical validity.  

One of the main virtues of this form is 
its model independence, serving as a global form able to cover reasonably 
large areas of dark energy phase space, in particular both classes of 
physics discussed in \S\ref{sec:dyn} -- thawing and freezing behavior. 
We can imagine, however, that future data will allow us to zero in on 
a particular region of phase space, i.e.\ area of the $w$-$w'$ plane, 
as reflecting the physical origin of acceleration.  In this case, 
we would examine more restricted, local parametrizations in an effort 
to distinguish more finely between physics models. 

\begin{figure}[!htb]
\begin{center}
\psfig{file=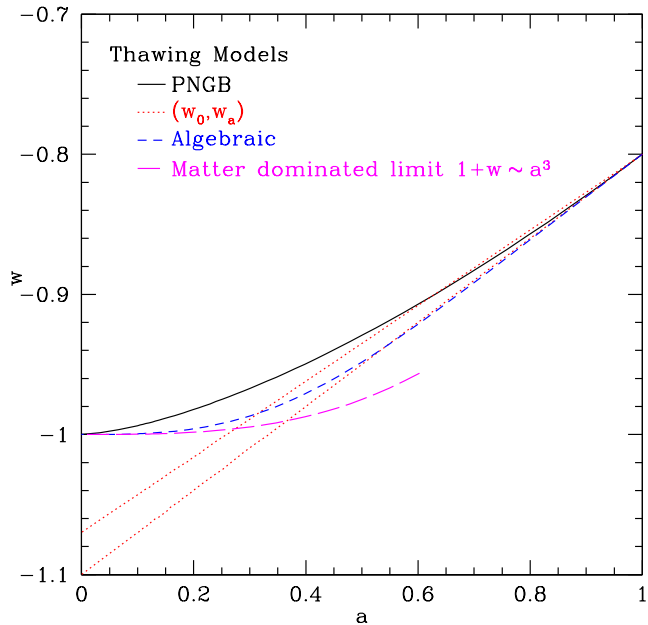,width=3.5in}
\caption{Within the thawing class of physics models one can hope to 
distinguish specific physics origins.  A canonical scalar field evolves 
along the early matter dominated universe behavior shown by the ``matter 
limit'' curve, then deviates as dark energy begins to dominate.  Such 
thawing behavior is well fit by the algebraic model.  A pseudoscalar 
field (PNGB) evolves differently.  Either can be moderately well fit 
by the phenomenological $(w_0,w_a)$ parametrization for the recent 
universe. 
}
\label{fig:thawmodels}
\end{center}
\end{figure}

First consider thawing models.  One well-motivated example is a 
pseudoscalar field \citep{FrieWaga}, a pseudo-Nambu Goldstone boson (PNGB), 
which can be well approximated by 
\beqa 
w'&=&F(1+w) \\ 
w(a)&=&-1+(1+w_0)\,a^F, 
\eeqa 
where $F$ is inversely proportional to the PNGB symmetry breaking energy scale 
$f$.  Scalar fields, however, thawing in a matter dominated universe, must at 
early times evolve along the phase space track $w'=3(1+w)$ \citep{CaldLin}, 
where $w$ departs from $-1$ by a term proportional to $a^3$.  One model 
tying this early required behavior to late times and building on the field 
space parametrization of \citet{CrittThaw} is the algebraic model 
of \citet{LinGRG}, 
\beq 
1+w=(1+w_0)\,a^p\,\left(\frac{1+b}{1+ba^{-3}}\right)^{1-p/3}, \label{eq:alg}
\eeq 
with parameters $w_0$, $p$ ($b=0.3$ is fixed).  Figure~\ref{fig:thawmodels} 
illustrates these different behaviors and shows matching by the global 
$(w_0,w_a)$ model, an excellent approximation to $w(z)$ out to $z\gs1$. 
More importantly, it reproduces distances to all redshifts to better than 
0.05\%.  
The key point, demonstrated in \citet{LinGRG}, is that use of any particular 
one of these parametrizations does not bias the main physics conclusions 
when testing consistency with the cosmological constant or the presence of 
dynamics.  Thus one can avoid a parametrization-induced systematic.

For freezing models, we can consider the extreme of the early dark 
energy model of \citet{DorRob}, with 3\% contribution to the energy 
density at recombination, near the upper limit allowed by current data 
\citep{DoranRW}. 
This model is specifically designed to represent dark energy that scales 
as matter at early times, transitioning to a strongly negative equation of 
state at later times.  If future data localizes the dark energy properties 
to the freezing region of phase space, one could compare physical origins 
from this model (due to dilaton fields) with, say, the $H^\alpha$ 
phenomenological model of \citet{DvaliTur} inspired by extra dimensions. 
Again a key point is that the global $(w_0,w_a)$ parametrization does 
not bias the physical conclusions, matching even the specialized, extreme 
early dark energy model to better than 0.02\% in distance out to 
$z\approx2$. 

Parametrizations to be cautious about are those that arbitrarily 
assume a fixed high redshift behavior, often setting $w=-1$ above some 
redshift.  These can strongly bias the physics \citep{dePLinPCA,wcon}. 
As discussed in \S\ref{sec:eff}, interesting and important physics 
clues may reside in the early expansion behavior.  

Bias can also ensue 
by assuming a particular functional form for the distance or Hubble 
parameter \citep{Jonsson}.  Even when a form is not assumed a priori, 
differentiating imperfect data (e.g.\ to derive the equation of state) 
leads to instabilities in the reconstruction \citep{Teg02,Recon}.  To 
get around this, one can attempt to smooth the data, but this returns 
to the problems of assuming a particular form and in fact can remove 
crucial physical information.  While the expansion history is innately 
smooth (also see \S\ref{sec:vern}), extraction of cosmological 
parameters involves differences between histories, which can 
have sharper features.

\subsection{Mirage of Lambda} \label{sec:mirage} 

As mentioned in \S\ref{sec:cmb}, interpreting the data without fully 
accounting for the possibility of dynamics can bias the theoretical 
interpretation.  We highlight here the phenomenon of the ``mirage of 
$\Lambda$'', where data incapable of precisely measuring the time 
variation of the equation of state, i.e.\ with the quality expected 
in the next five years, can nevertheless apparently indicate with 
high precision that $w=-1$. 

\begin{figure}[!htb]
\begin{center}
\psfig{file=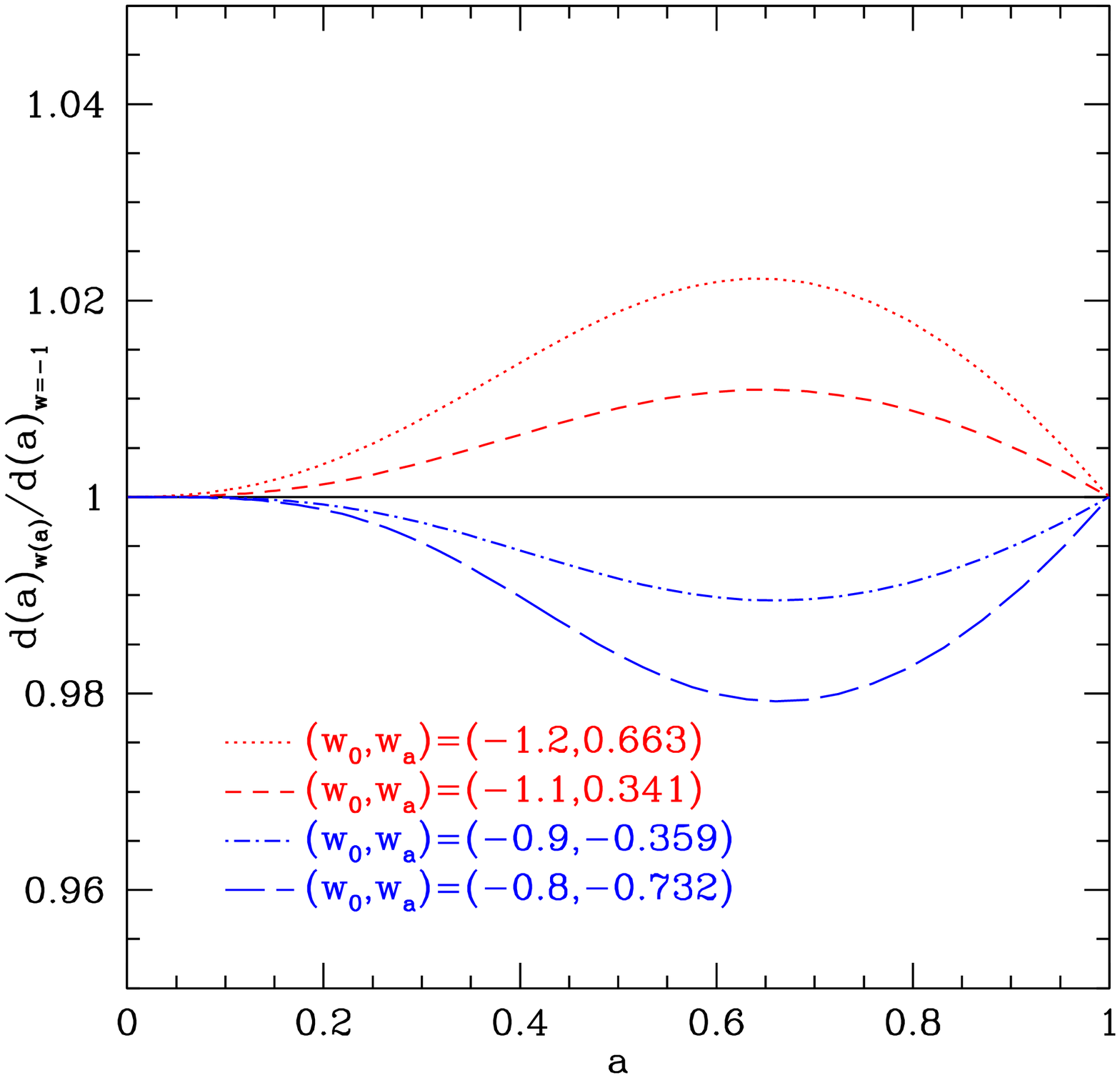,width=3.in}
\psfig{file=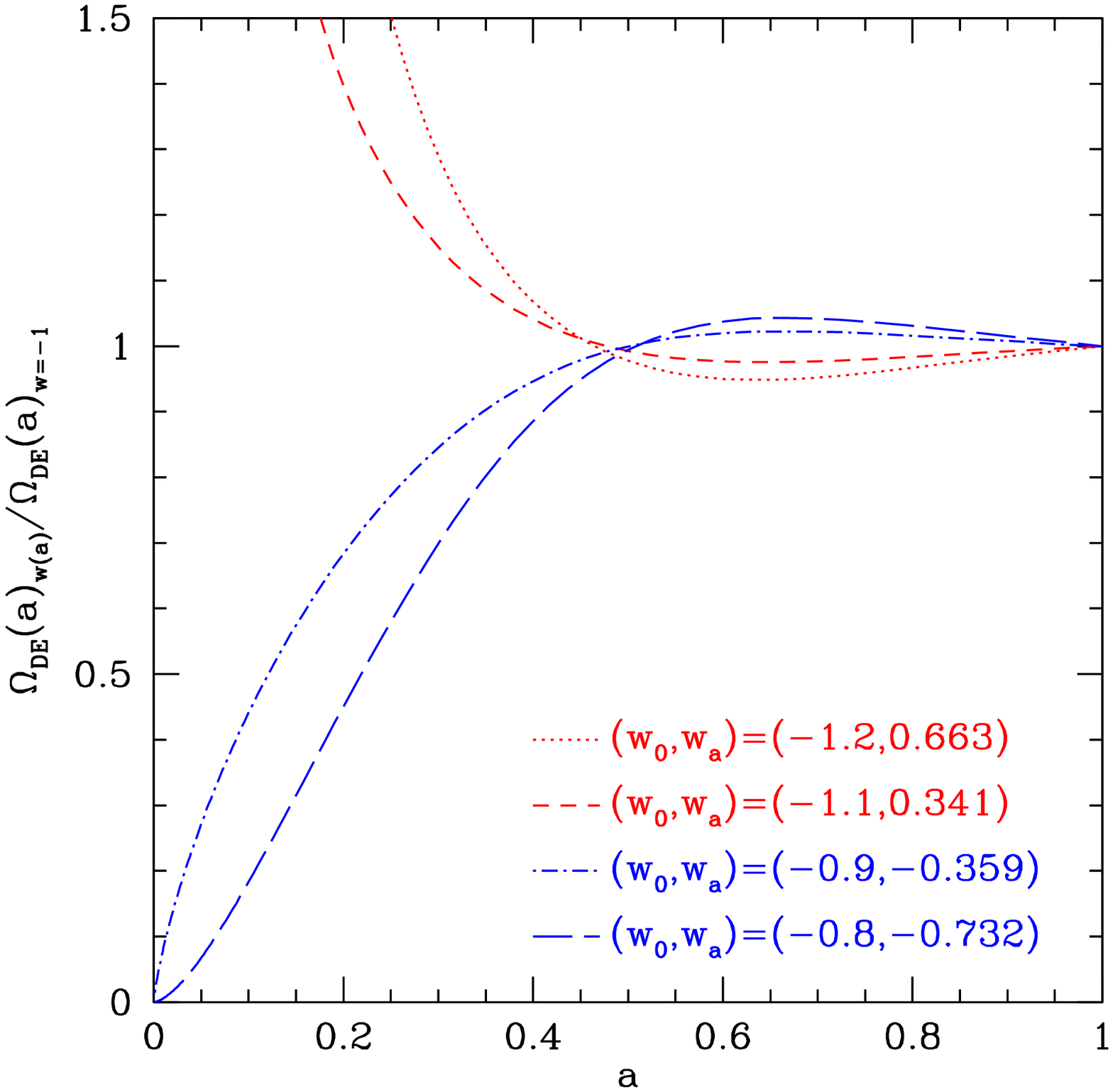,width=3.in}
\psfig{file=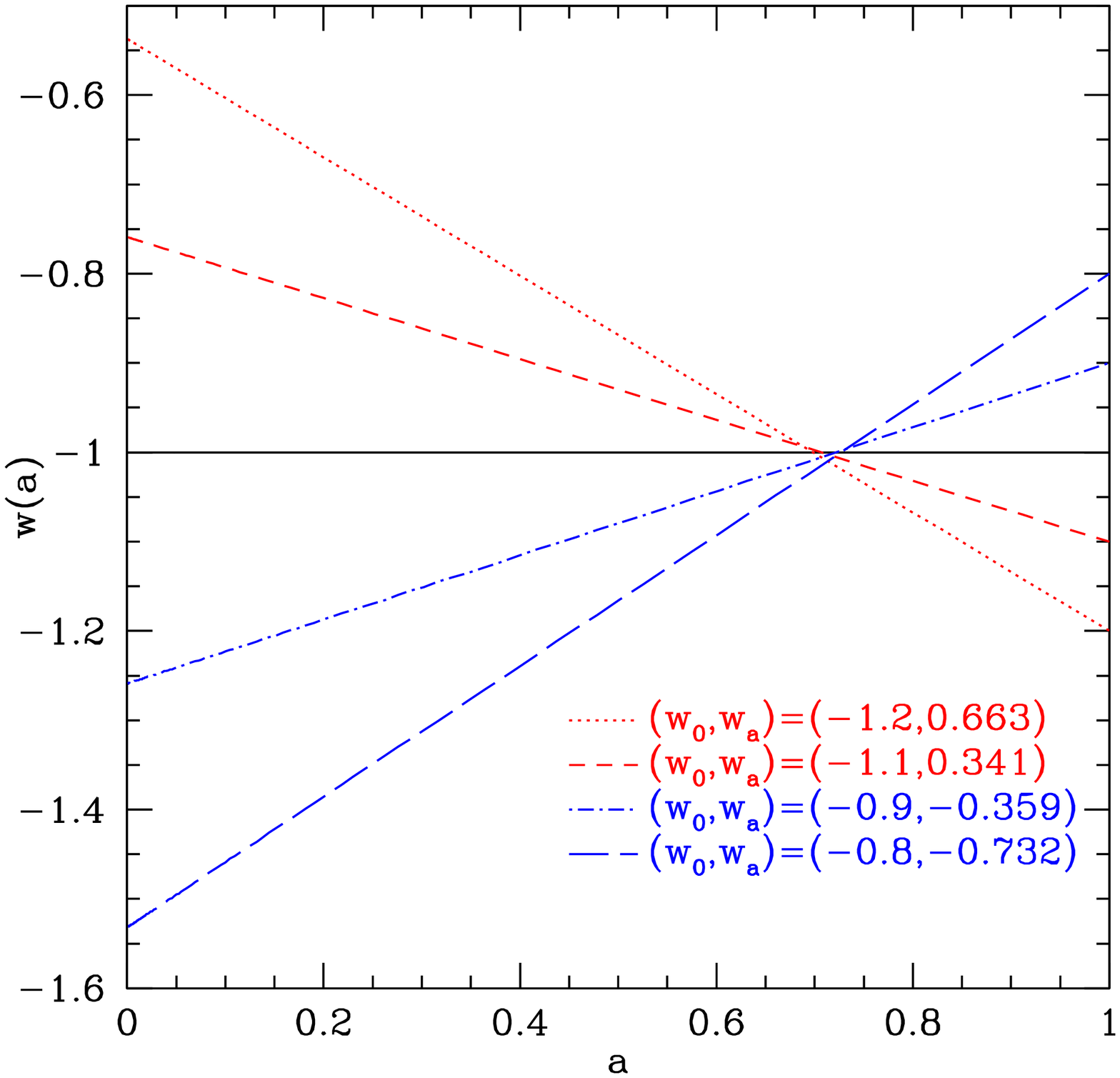,width=3.in}
\caption{Matching the distance to CMB last scattering between dark energy 
models leads to convergence and crossover behaviors in other cosmological 
quantities.  The top left panel illustrates the convergence in the 
distance-redshift relation, relative to the $\Lambda$CDM case, for models 
with $w_0$ ranging from $-0.8$ to $-1.2$ and corresponding time variation 
$w_a$.  The top right panel illustrates the related convergence and 
crossover in the dark energy density $\Omega_{\rm DE}(a)$, and the bottom 
panel shows how the CMB matching necessarily leads to a crossover with 
$w=-1$ at the key redshift for sensitivity of low redshift experiments.  
This crossover in $w(z)$ leads to the mirage of $\Lambda$, and is impelled 
by the physics not the functional form. 
}
\label{fig:miraged}
\end{center}
\end{figure}

Suppose the distance to CMB last scattering -- an integral measure of 
the equation of state -- matches that of a $\Lambda$CDM model.  Then 
under a range of circumstances, low redshift ($z\la1$) measurements 
of distances can appear to show that the equation of state is within 5\% 
of the cosmological constant value, even in the presence of time variation.  
Figure~\ref{fig:miraged} illustrates 
the sequence of physical relations leading to this mirage.

Matching the distance to last scattering imposes a relation between the 
cosmological parameters, here shown as a family of $(w_0,w_a)$ models 
holding other parameters fixed.  The convergence in their distances 
beginning at $a\approx0.65$ is associated with a similar convergence 
in the fractional dark energy density, and a matching in $w(z)$ at 
$a\approx0.7$.  Note that even models with substantial time variation, 
$w_a\approx1$, are forced to have $w(z\approx0.4)=-1$, i.e.\ look like 
the cosmological constant.  This is dictated by the innate cosmological 
dependences of distance and is robust to different parameter choices; 
see \citet{wcon} for more details.  (Note the matching in $\Omega_{\rm 
DE}(a)$ forced at $a\approx0.5$ has implications for the linear growth 
factor and nonlinear power spectrum, as explored in \citet{FrancisLL}.) 

\begin{figure}[!htb]
\begin{center}
\psfig{file=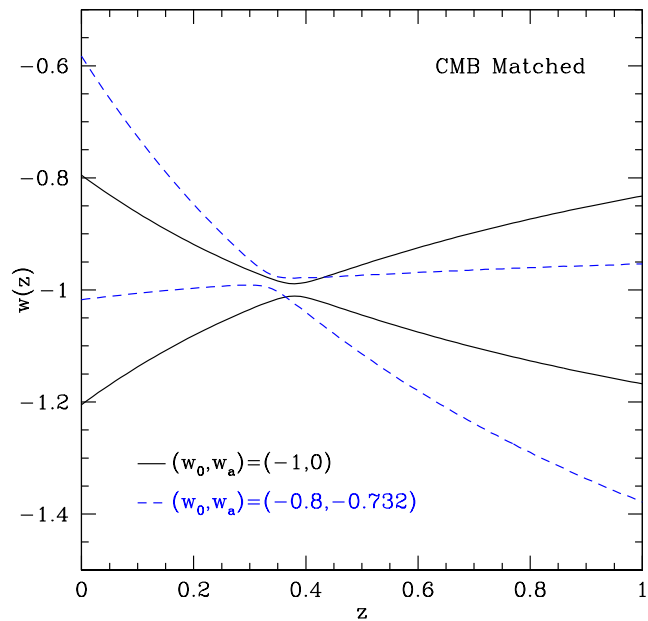,width=3.5in}
\caption{If CMB data are consistent with $\Lambda$CDM, this can create 
a mirage of $\Lambda$ for lower redshift distance data even if the 
dark energy has substantial time variation.  The curves show simulated 
68\% confidence regions for $w(z)$ for two different CMB-matched models. 
The value of the equation of state $w(z)=-1$ necessarily, for each one 
at a redshift close to the ``sweet spot'' 
or pivot redshift.  Experiments insufficiently precise to see time 
variation will think $w=-1$ to high precision (the width of the narrow 
waist at $z\approx0.38$, here 1\%) even if the true behavior is 
drastically different. 
}
\label{fig:miragew}
\end{center}
\end{figure}

To see beyond the mirage of $\Lambda$, or test its reality, requires 
measurements capable of directly probing the time variation with 
significant sensitivity (hence extending out to $z\approx1.7$ as shown in 
\S\ref{sec:distgen}).  Current and near term experiments that may 
show $w\approx-1$ to a few percent can induce a false sense of security 
in $\Lambda$.  In particular, the situation is exacerbated by the pivot 
or decorrelation redshift (where the equation of state is measured most 
precisely) of such experiments probing to $z\approx1$ being close to 
the matching redshift imposed on $w(z)$; so, given CMB data consistent 
with $\Lambda$, such experiments will measure $w=-1$ with great 
precision, but possibly quite inaccurately.  See Figure~\ref{fig:miragew} 
for a simulation of the possible data interpretation in terms of the 
mirage of $\Lambda$ from an experiment capable of achieving 1\% minimum 
uncertainty on $w$ (i.e.\ $w(z_{\rm pivot})$).  Clearly, the time 
variation $w_a$ is the important physical quantity allowing us to see 
through the mirage, and the key science requirement for next generation 
experiments.

\subsection{Inhomogeneous data sets} 

Turning from theory to data analysis, another source of systematics that 
can lead to improper cosmological 
conclusions are heterogeneous data sets.  This even holds with data all 
for a single cosmological probe, e.g.\ distances.  Piecing together 
distances measured with different instruments under different conditions 
or from different source samples opens the possibilities of 
miscalibrations, or offsets, between the data. 

While certainly an issue when combining, say, supernova distances with 
baryon acoustic oscillation distances, or gravitational wave siren 
distances, we illustrate the implications even for heterogeneous 
supernova samples.  An offset between the magnitude calibrations can 
have drastic effects on cosmological estimation (see, e.g., 
\citet{LinBias}).  For example, very low redshift ($z<0.1$) supernovae 
are generally observed with very different telescopes and surveys than 
higher redshift ($z>0.1$) ones.  
Since the distances in the necessary high redshift ($z\approx1-1.7$) sample 
require near infrared observations from space then the crosscalibration 
with the local sample (which requires very wide fields and more rapid 
exposures) requires care.  The situation is exacerbated if the space sample 
does not extend down to near $z\approx0.1-0.3$ and a third data set 
intervenes.  This gives a second crosscalibration needed to high accuracy. 

\begin{figure}[!htb]
\begin{center}
\psfig{file=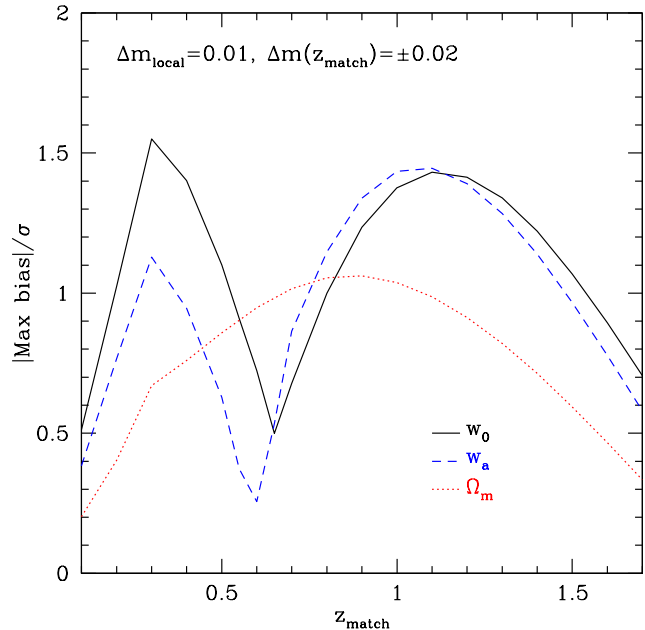,width=3.5in}
\caption{Heterogeneous datasets open issues of imperfect crosscalibration, 
modeled here as magnitude offsets $\Delta m$.  One scenario involves 
calibration between a local ($z<0.1$) spectroscopic set and a uniform 
survey extending from $z\approx0.1-1.7$.  This imposes cosmological 
parameter biases given by the intersection of the curves with the left 
axis.  Another scenario takes the high redshift data to consist of two, 
heterogeneous sets with an additional offset $\Delta m$ at some 
intermediate matching redshift $z_{\rm match}$.  (When $z_{\rm match}=0.1$ 
this corresponds to the first scenario with no extra offset.) 
}
\label{fig:dmoff}
\end{center}
\end{figure}

Figure~\ref{fig:dmoff} demonstrates the impact on cosmology, with the 
magnitude offset leading to bias in parameter estimation.  If there is 
only a local distance set and a homogeneous space set extending from low 
to high redshift, with crosscalibration at the 0.01 mag level, then the 
biases take on the values at the left axis of the plot: a small fraction 
of the statistical dispersion.  However, with an intermediate data set, 
the additional heterogeneity from matching at some further redshift 
$z_{\rm match}$ (with the systematic taken 
to be at the 0.02 mag level to match a ground based,  non-spectroscopic 
experiment to a space based spectroscopic experiment) runs the risk of 
bias in at least one parameter by of order $1\sigma$.  Thus cosmological 
accuracy advocates as homogeneous a data set as possible, ideally from 
a single survey.

Similar heterogeneity and bias can occur in baryon acoustic oscillation 
surveys mapping the expansion when the selection function of galaxies 
varies with redshift.  If the power spectrum shifts between samples, due 
for example to different galaxy-matter bias factors between types of 
galaxies or over redshift, then calibration 
offsets in the acoustic scale lead to biases in the cosmological parameters. 
Again, innate cosmology informs survey design, quantitatively determining 
that a homogeneous data set over the redshift range is advantageous.

\subsection{Miscalibrated standard} 

Miscalibration involving the basic standard, i.e.\ candle luminosity or 
ruler scale, has a pernicious effect biasing the expansion history mapping. 
This time we illustrate the point with baryon acoustic oscillations.  If 
the sound horizon $s$ is improperly calibrated, with an offset $\delta s$ 
(for example through early dark energy effects in the prerecombination 
epoch \citep{DoranLSS,LinRob}), then every baryon acoustic oscillation 
scale measurement $\tilde d(z)=d(z)/s$ and $\tilde H(z)=sH(z)$ will be 
miscalibrated. 
Due to the redshift dependence of the untilded quantities, the offset will 
vary with redshift, looking like an evolution that can be confused with a 
cosmological model biased from the reality. 

To avoid this pitfall, analysis must include a calibration parameter for 
the sound horizon (since CMB data does {\it not\/} uniquely determine it 
\citep{EisWh,LinRob}), in exact analogy to the absolute luminosity 
calibration parameter ${\mathcal M}$ required for supernovae.  That is, 
the standard ruler must be 
standardized; {\it assuming\/} standard CDM prerecombination for the 
early expansion history blinds the analysis to the risk of biased 
cosmology results. 

The necessary presence of a standard ruler calibration parameter, call it 
${\mathcal S}$, leads to an increase in the $w_0$-$w_a$ contour area, and 
equivalent decrease in the ``figure of merit'' defined by that area, by 
a factor 2.3.  Since we 
do not know a priori whether the high redshift universe is conventional CDM 
(e.g.\ negligible early dark energy or coupling), 
neglecting ${\mathcal S}$ for BAO is as improper as neglecting 
${\mathcal M}$ for supernova standard candle calibration.  (Without the 
need to fit for the low redshift calibration ${\mathcal M}$, SN would enjoy 
an improvement in ``figure of merit'' by a factor 1.9, similar to the 2.3 
that BAO is given when neglecting the high redshift calibration 
${\mathcal S}$.)  

For supernovae, in addition to the fundamental calibration of the absolute 
luminosity, experiments must tightly constrain any evolution in the 
luminosity \citep{Coping}.  This requires broadband flux data from soon 
after explosion to well into the decline phase, and spectral data over a 
wide frequency range.  Variations in supernova properties that do not 
affect the corrected peak magnitude do not affect the cosmological 
determination.

\subsection{Malmquist bias} 

Distance measurements from the cosmological inverse square law of flux 
must avoid threshold effects where only the brightest sources at a given 
distance can be 
detected, known as Malmquist bias.  Suppose the most distant, and hence 
dimmest, sources were close to the detection threshold.  We can treat 
this selection effect as a shift in the mean magnitude with a toy model, 
\beq 
\Delta m=-A\,\frac{z-z_\star}{0.95-z_\star},\qquad z>z_\star\,.
\eeq 
This then propagates into a bias on the cosmological parameter fit to 
the data.  

Consider a data set of some 1000 supernovae from $z=0-1$, 
with the Malmquist bias setting in at $z_\star=0.8$ (where ground based 
spectroscopy begins to get quite time intensive and many spectral features 
move into the near infrared).  The bias in a cosmological parameter 
relative to its uncertainty is then 
\beq 
\frac{\delta p}{\sigma_p}=(1.5,1.0,1.2)\times\frac{A}{0.1\ {\rm mag}}
\eeq 
for $p=(\om,w_0,w_a)$.  Thus, the Malmquist bias must be limited to less 
than 0.05 mag at $z=0.95$ to prevent significant bias in the derived 
cosmological model.  In fact, this is not a problem for 
a well designed supernova survey since the requirement of mapping out 
the premaximum phase of the supernova ensures sensitivity to fluxes 
at least two magnitudes below detection threshold.

\subsection{Other issues} 

In addition to the theory interpretation and data analysis systematics 
discussed in this section, recall the fundamental theory systematics of 
\S\ref{sec:gravsys} and Table~\ref{tab:thysys}.  We finish with 
a very brief mention of some other selected data and data analysis 
systematics issues of importance that are often underappreciated and 
that must be kept in mind for proper survey design and analysis.  

\begin{figure}[!htb]
\begin{center}
\psfig{file=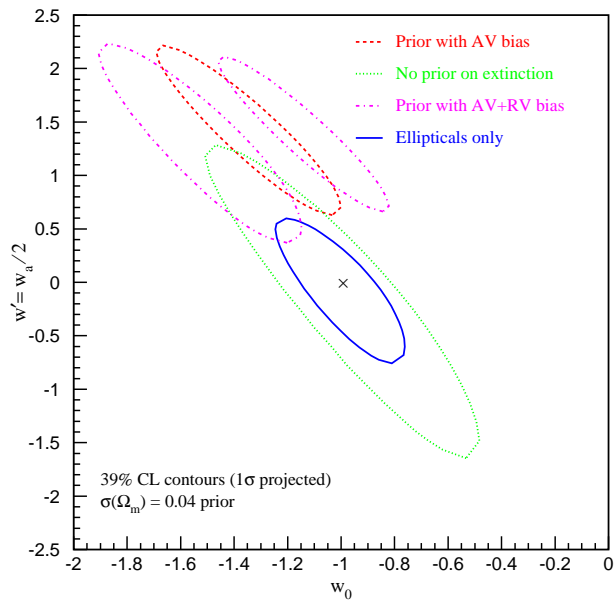,width=3.5in}
\caption{Assuming a prior on dust properties ($A_V$ and $R_V$) in order 
to reduce extinction errors can cause systematic deviations in the 
magnitudes.  These give strong biases to the equation of state, leading 
to a false impression of a transition from $w<-1$ to $w>-1$.  To avoid 
this systematic, one can use samples with minimal extinction (elliptical 
galaxy hosted supernovae) or obtain precise multiwavelength data that 
allows for fitting the dust and color properties.  Based on \citet{LinMiq}. 
}
\label{fig:dustprior}
\end{center}
\end{figure}

\begin{itemize} 

\item {\it Sample variance:} Along a given line of sight, the local 
distance measures anchoring the Hubble diagram can be influenced by 
coherent velocity flows, throwing off the derived cosmology 
\citep{HuiGreene,CooCalVel,ConleyVel}. 
The local distances should therefore be well into the Hubble flow and 
the sources distributed widely on the sky. 
In addition, the mass distribution along the line of sight may not be 
representative of the homogeneous model and gravitational lensing can 
lead to coherent magnification effects (relevant for standardized 
candles) and alterations of the measured three dimensional clustering 
(important for baryon acoustic oscillations).  See, e.g., 
\citet{Pencil,LoVerdeHui}. 
For these reasons and others, ``pencil beam'' surveys can be fraught with 
systematics and are poor survey design. 

\item {Analytic marginalization:} The calibration parameter, e.g.\ 
${\mathcal M}$ combining the absolute luminosity and Hubble constant in 
the case of supernovae, is often referred to as a nuisance parameter 
but its proper treatment is essential.  Although in some $\chi^2$ 
formulas for the distance-redshift relation it is not written explicitly, 
it is implicit and cannot be ignored.  More subtle is the issue of analytic 
marginalization over it -- this must be used with great care 
as the distribution of ${\mathcal M}$ is actually 
non-Gaussian due to interaction with other supernovae peak magnitude 
fitting quantities (such as the lightcurve width and color terms) 
\citep{Kowal}. 
Further subtleties exist between marginalization and minimization in 
a multidimensional fit space \citep{SteinhMinim,ConleyVel}, and most 
analysis from raw data to quoted parameters actually employs minimization 
techniques. 

\item {\it Extinction priors:} Since the dimming and reddening due to 
dust effects on supernovae are one-sided (i.e.\ dust does not increase 
the flux), they are highly non-Gaussian and must be treated with care. 
Any deviation between an assumed prior for extinction and the truth, that 
is not constant in redshift, can bias the cosmology results.  See 
Figure~\ref{fig:dustprior} and the handy systematics calculator SMock 
\citep{Smock} for examples.  Several analysis 
techniques avoid this pitfall by fitting for dust and intrinsic color 
globally, without assuming a prior, though this requires high quality 
data over several wavelength bands. 

\end{itemize}

\section{Future prospects} \label{sec:fut} 

The main uncertainty in our future ability to map the cosmological 
expansion is the level of control of systematics we can achieve.  
In a reductio ad absurdum we can say that since one supernova explodes 
every second in the universe we might measure $10^7$ per year, giving 
distance accuracies of $10\%/\sqrt{10^7}=0.003\%$ in ten redshift 
slices over a ten year survey; or counting every acoustic mode in 
the universe per redshift slice spanning the acoustic scale $\lambda$ 
determines the power spectrum to $\lambda^3/(H^{-2}\lambda)=(1/30)^2$ 
or 0.1\%; or measuring weak lensing shear across the full sky with 
0.1'' resolution takes individual 1\% shears to the 0.0003\% level. 
Statistics is not the issue: understanding of systematics is. 

\subsection{Data and systematics} 

However it is much easier to predict the future of statistical 
measurements than systematic uncertainties.  It is difficult to 
estimate what the future prospects really are.  Large surveys are 
being planned assuming that uncertainties will be solved -- or if the 
data cannot be used for accurately mapping the cosmic expansion it 
will still prove a cornucopia to many fields of astrophysics. 
Moreover, an abundance of surveys mentioned in the literature 
are various levels of vaporware: many have never passed a national 
peer review or been awarded substantial development funding or had 
their costs reliably determined.  Rather than our listing such 
possibilities, the reader can peruse national panel reports such 
as the ESA Cosmic Vision program \citep{CosmicVision}, US National 
Academy of Sciences' Beyond Einstein Program Assessment Committee 
report \citep{bepac}, or the upcoming US Decadal Survey of Astronomy 
and Astrophysics \citep{nextdecadal}. 

The current situation regarding treatment of systematics is mixed.  In 
supernova cosmology, rigorous identification and analysis of systematics 
and their effects on parameter estimation is (almost) standard 
procedure.  In weak lensing there exists the Shear Testing Programme 
\citep{step}, producing and analyzing community data challenges.  Other 
techniques have less organized systematics analysis, where it exists, 
and the crucial rigorous comparison of independent data sets (as in 
\citet{Kowal}) is rare. 
Control methods such as blind analysis are also rare.  We cannot yet 
say where the reality will lie between the unbridled statistical optimism 
alluded to in the opening of this section and current, quite modest 
measurements.  Future prospects may be bright but considerable 
effort is still required to realize them.

\subsection{Mapping resolution} \label{sec:vern}

Since prediction of systematics control is difficult, let us turn instead 
to intrinsic limits on our ability to map the expansion history -- limits 
that are innate to cosmological observables.  The expansion history $a(t)$, 
like distances, is an integral over $H=d\ln a/dt$, and does not respond 
instantaneously to the evolution of energy density (which in turn is an 
integral over the equation of state).  As seen in 
Figures~\ref{fig:sensd}-\ref{fig:sensh}, the cosmological kernel for 
the observables is broad -- no fine toothed comb exists for studying the 
expansion.  This holds even where models have rapid time variation $w'>1$, 
as in the e-fold transition model, or when considering principal components 
(see \citet{Vernier,dePLinPCA} for illustrations).  

(This resolution limit 
on mapping the expansion is not unique to distances -- the growth factor 
is also an integral measure and the broad kernel of techniques like weak 
lensing is well known.  The Hubble parameter determined through BAO, say, 
requires a redshift shell thickness $\Delta z\gs0.2$ to obtain sufficient 
wave modes for good precision, limiting the mapping resolution.) 

Thus, due to the innate cosmological dependences of observables, plus 
additional effects such as Nyquist and statistics limits of wavemodes in 
a redshift interval and coherence of systematics over redshift 
\citep{Vernier}, we cannot expect mapping of the cosmological expansion with 
finer resolution than $\Delta z\approx0.2$ from next generation data.

\subsection{Limits on cosmic doomsday} \label{sec:doom}

Finally, we turn from the expansion history to the expansion future. 
As pointed out in \S\ref{sec:fate}, to determine the fate of our universe 
we must not only map the past expansion but understand the nature of the 
acceleration before we can know the universe's destiny -- eternal 
acceleration, fading of dark energy, or recollapse. 

\begin{figure}[!htb]
\begin{center}
\psfig{file=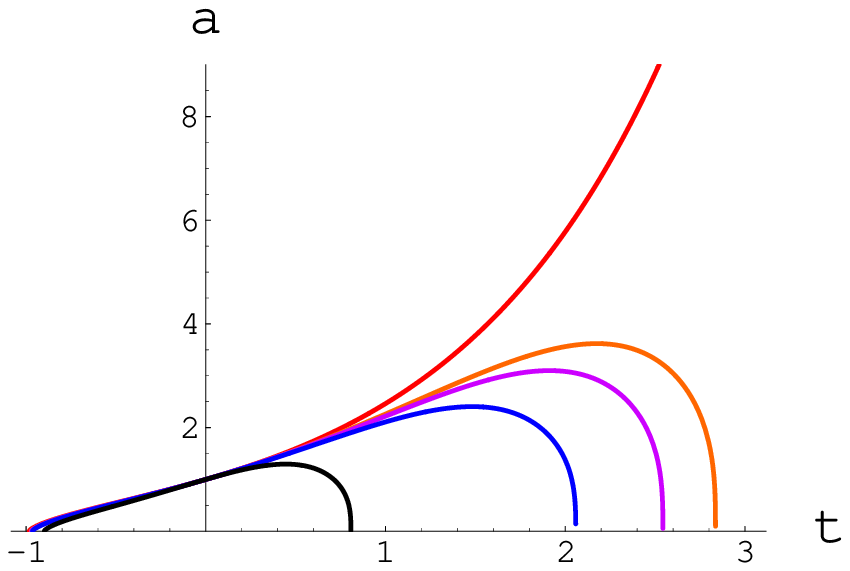,width=3.in}
\psfig{file=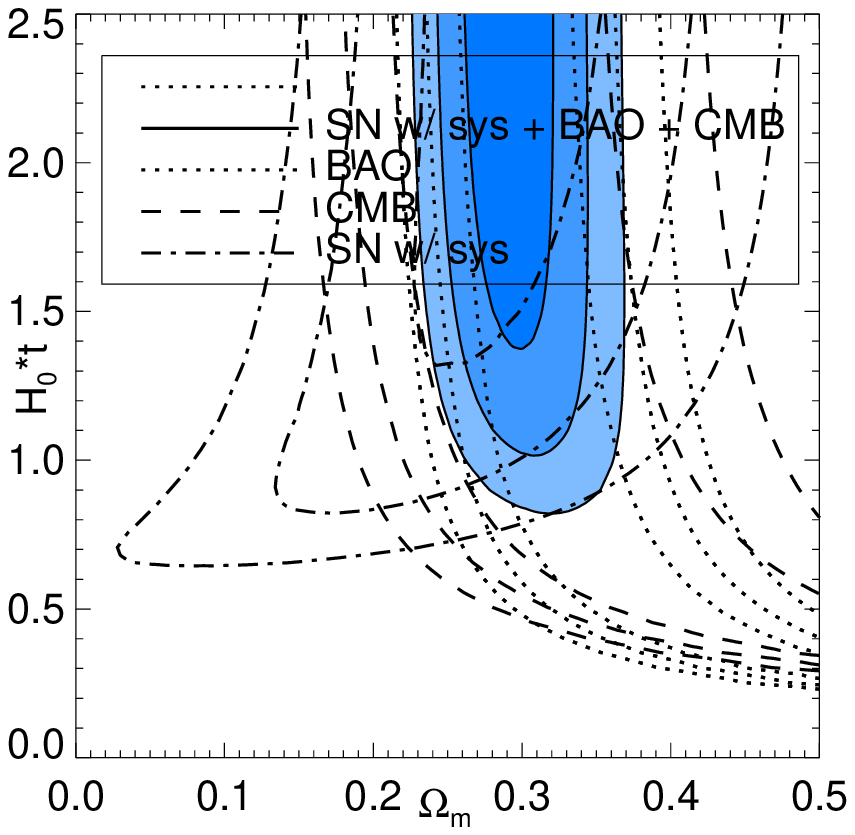,width=3.in}
\caption{The future expansion history in the linear potential model 
has a recollapse, or cosmic doomsday.  {\it Left panel:} Precision 
measurements of the 
past history are required to distinguish the future fate, with the 
five curves corresponding to five values for the potential slope 
(the uppermost curve is for a flat potential, i.e.\ a cosmological 
constant).  From \citet{Kratoch}.  {\it Right panel:} Current data 
constraints estimate cosmic doomsday will occur no sooner than 
$\sim$1.5 Hubble times from now.  From \citet{Rubin}.
}
\label{fig:doomsday}
\end{center}
\end{figure}

We do not yet have that understanding, but suppose we assume that the linear 
potential model of dark energy \citep{Linde86}, perhaps the simplest 
alternative to a cosmological constant, is correct.  Then we can estimate 
the time remaining before the fate of recollapse: the limit on cosmic 
doomsday.  The linear potential model effectively has a single equation 
of state parameter and is well approximated by the family with 
$w_a=-1.5(1+w_0)$ when $w_0$ is not too far from $-1$.  Curves of the 
expansion history (and future) are illustrated in the left panel of 
Figure~\ref{fig:doomsday}; the doomsday time from the present is 
given by 
\beq 
t_{\rm doom}\approx 0.5\,H_0^{-1}(1+w_0)^{-0.8}. 
\eeq 

The current best constraints from data (cf.\ \citet{Kratoch} for 
future limits) appear in the right panel of Figure~\ref{fig:doomsday}, 
corresponding to $t_{\rm doom}>24$ Gy at 68\% confidence level. 
It is obviously of interest to us to know whether the universe will 
collapse and how long until it does, so accurate determination of $w_a$ 
is important!  The difference between doomsday in only two Hubble times 
from now and in three (a whole extra Hubble time!) is only a difference 
of 0.12 in $w_a$.

\section{Conclusions} \label{sec:concl} 

Mapping the cosmological expansion is a key endeavour in our quest for 
understanding physics -- gravitation and other forces, spacetime, the 
vacuum -- and the origin, evolution, and future of the universe.  The 
Big Bang model of a hot, dense, early universe expanding and adiabatically 
cooling, and forming structure through gravitational instability from 
primordial seed perturbations, is remarkably successful and simple.  
The concordance cosmology is (close to) spatially flat and 13.7 billion 
years old.  This clear statement represents a substantial advance of 
our knowledge over a decade ago. 

The discovery of the acceleration of the cosmic expansion created a 
renaissance of cosmological exploration, offering the hope of connecting 
quantum physics and gravitation, extra dimensions and the nature of 
spacetime, while severing the bonds between geometry and destiny.  This 
opens up completely two premier questions in science: the origin and the 
fate of the universe.  To understand the nature of the gravitationally 
repulsive dark energy pulling the universe apart, we must map the 
cosmological expansion in greater detail and accuracy than ever envisioned. 

We have shown that a considerable part of the optimal approach for mapping 
the expansion history is set purely by the innate cosmological dependences. 
Observational programs must follow these foundations as basic science 
requirements: in particular the need for a wide range of redshifts, 
$z\approx0-2$ and robust anchoring to either low or high redshift. 
Our capabilities for measuring the expansion through direct geometric 
probes are increasing, and techniques are continuing to develop. 
There are no short cuts -- detailed design of successful surveys works 
within this framework with the purpose to minimize systematic uncertainties 
in the measurements.  

Every technique has systematic uncertainties, appearing in multiple guises. 
While observational systematics are most familiar, arising even in purely 
geometric techniques, equally important are issues in the data analysis, 
e.g.\ combining heterogeneous data sets, and in the theoretical 
interpretation, e.g.\ susceptibility to biases from assumptions of high 
redshift behavior or of non-expansion physics (perturbation growth 
behavior, coupling, etc.). 

We have given several concrete examples of the effects of systematic 
uncertainties for various probes.  These provide a cautionary tale for 
survey design, as we are now entered into the systematics dominated data 
era -- and may well soon approach the theory systematics era.  We cannot 
rely on assumptions that any part of the dark sector is simple and ignorable 
while we concentrate on another aspect.  For true progress, we emphasized 
the role of complementarity 
in building from robust, clean answers to more complex investigations. 
Probes employing growth of structure give windows on both expansion per se 
and gravitational laws, and we commented that the excitement of testing 
general relativity 
is equally matched with the challenge of creating a framework for analysis. 

Acceleration of the cosmic expansion heralds a revolution in physics, if 
we can characterize and understand it.  To comprehend this new aspect of 
the universe, we must map the expansion not only at recent epochs, but 
encompass the early universe.  Once we have garnered sufficient 
understanding, the prize is answering the question of the fate of the 
universe, now unbound from the question of the cosmic geometry.  The 
good news is that we likely have at least 24 billion years to do so!

\ack

I thank the Aspen Center for Physics for a superb working environment, 
and Los Alamos National Laboratory and the Santa Fe 07 workshop, 
University of Heidelberg, and the 
Dark Cosmology Centre and Niels Bohr Summer Institute, for hospitality. 
I gratefully acknowledge contributions by Georg Robbers, Ramon Miquel, 
and David Rubin of Figures~\ref{fig:dlsss}, \ref{fig:dustprior}, and 
\ref{fig:doomsday} respectively.  Colleagues too numerous to mention 
helped form my thinking on this wide ranging topic, but I will single 
out Bob Wagoner for laying the foundations. 
This work has been supported in part 
by the Director, Office of Science, Department of Energy under grant 
DE-AC02-05CH11231.

\end{document}